\documentclass[12pt]{iopart}
\input epsf

\newcommand{\gl}[1]{Eq. (\ref{#1})}
\newcommand{\gls}[2]{Eqs. (\ref{#1}) and (\ref{#2})}
\newcommand{\beq}[1]{\begin{equation}\label{#1}}
\newcommand{\eeq}{\end{equation}}

\newcommand{\bsq}[1]{\begin{subequations}\label{#1}}
\newcommand{\esq}{\end{subequations}}
\newcommand{\beqa}[1]{\begin{eqnarray}\label{#1}\nonumber}
\newcommand{\eeqa}{\end{eqnarray}}
\newcommand{\fur}{\qquad\mbox{for }\, }
\newcommand{\wer}{\qquad\mbox{where }\quad}

\newcommand{\vek}[1]{{\bf #1}}

\begin{document}

\topical{Structure of Colloid-Polymer Suspensions}

\author{Matthias Fuchs\footnote{permanent address: Physik-Department,
Technische Universit\"at M\"unchen, 85747 Garching, Germany}\\}
\address{
Department of Physics and Astronomy, The University of Edinburgh,
JCMB King's Buildings, Edinburgh EH9 3JZ, United Kingdom\\
}

\author{Kenneth S. Schweizer}
\address{Departments of Materials Science and Engineering and
Chemistry, and Materials Research Laboratory, University of
Illinois, Urbana, Illinois 61801, USA\\\bigskip }
\date{\today}

\begin{abstract}
We discuss structural correlations in mixtures of free polymer and
colloidal particles based on a microscopic, 2-component liquid
state integral equation theory. Whereas in the case of polymers
much smaller than the spherical particles the relevant polymer
degree of freedom is the center of mass, for polymers larger than
the (nano-) particles conformational rearrangements need to be
considered. They have the important consequence that the polymer
depletion layer exhibits two widely different length scales, one
of the order of the particle radius, the other of the order of the
polymer radius or the polymer density screening length in dilute
or semidilute concentrations, respectively. Because we find a
spinodal instability (mostly) below the overlap concentration, the
latter length is (mostly) set by the radius of gyration. As a
consequence of the structure of the depletion layer, the
particle-particle correlations depend on both length scales for
large polymers. Because of the high local compressibility of large
polymers, the local depletion layer is a strong function of
particle density, but a weak function of polymer concentration.
The amplitude of the long-ranged tail of the depletion layer
asymptotically also depends only on colloid concentration, while
the range increases upon approaching the (mean-field) spinodal.
The colloid correlations may be understood as characteristic for
particles with a short ranged potential when small polymers are
added, and as characteristic for particles with a long-ranged, van
der Waals like attraction when the added free polymer coils are
much larger. Small polymers fill the voids between the particles
rather homogeneously, exhibiting correlations inside the mesh
(which gets squeezed by the colloids) and Porod-like correlations
for larger distances. The structure factor of large polymers,
however, exhibits no ramified mesh and  becomes a Lorentzian
characterized by the mixture correlation length, which diverges at
the spinodal.
\end{abstract}

\submitted{{\noindent \it }}

\section{Introduction}

Polymer-colloid suspensions are generally 3-component systems
composed of solid or impenetrable particles, macromolecules, and
solvent \cite{1,2,3}. The latter are usually small molecules compared to
the polymers coils and colloids, and thus solvent is generally not
explicitly treated by theories but rather taken as a background
continuum which can influence the effective interactions between
the other two components. The strong geometric asymmetry between
polymers and particles is at the heart of their novel physical
behavior and useful material properties. All size asymmetry
regimes are of importance, spanning the range from the traditional
problem of large colloids and small polymers, to small
nanoparticles and large macromolecules. The statistical geometry
of polymers can be highly variable including structures such as
ideal random coils, self-avoiding random walks, semiflexible
persistent chains, star-branched macromolecules, rigid rods, and
fractal aggregates. The steric interactions between polymers, and
polymers and colloids, depend on macromolecular geometry.
Moreover, if the polymers are sufficiently flexible, the presence
of colloids can potentially perturb their statistical
conformation.

Great diversity also characterizes the particles spanning the
range from the most common spherical geometry, to aspherical
viruses, to highly anisotropic colloidal rods and plate or
discotic type structures. The particles can be organic or
inorganic solids, dendrimers, surfactant micelles, globular
proteins, vesicles, or soft crosslinked microgels. Although the
surfaces of particles are often homogeneous, they need not be. For
example, both synthetic dendrimers and natural proteins possess
chemically heterogeneous, or patchy, surfaces. Repulsive excluded
volume forces are universally present, and the geometric diversity
provides a rich set of possibilities for packing in such
suspensions. Of course, other forces are generally present and can
be rationally tuned including van der Waals attractions, Coulombic
forces, and specific interactions such as the hydrophobic effect
\cite{1,2,3}.

There are also diverse mixture composition regimes which often are
of interest to different scientific and engineering communities.
In colloid or nanoparticle science, polymers are often used as low
concentration additives to manipulate the colloidal suspension
properties \cite{1,2,3}. On the other hand, in polymer science solid
particles are commonly employed as fillers to manipulate the
properties of dense polymer melts, glasses or rubbers \cite{4}, or as
additives to modify the behavior of polymer solutions which can
form soft gels \cite{5}. True composite materials where the volume
fraction of particles and polymers are comparable are of
significant scientific and applications interest.

In the present feature article, we focus on the simplest
realization of these systems: hard spheres and flexible polymer
chains under athermal ``good'' solvent conditions. This is the
most fundamental and generic system characterized solely by hard
core repulsive interactions between all species. The statistical
mechanical problem is a purely entropic "packing problem".  For
dilute and semidilute polymer concentrations \cite{6}, there are only
three fundamental dimensionless variables: (i) colloid packing
fraction, $\phi_{c}$ , (ii) polymer monomer or segment
concentration reduced by its value at the dilute-semidilute
crossover (where polymer-polymer interactions in a particle free
solution become important), $c/c^{*}$, and (iii) size asymmetry,
$R_{g}/R$, where $R$ is the colloid radius and $R_{g}$ the polymer
radius-of-gyration. Despite its simple model nature, the physical
behavior of such mixtures is rich and complex
\cite{1,2,3,7,8,9,10,11,12,13,14,15}.   In
equilibrium the homogeneous fluid phase competes with fluid-fluid
and fluid-crystal phase separated states. Nonequilibrium glass or
gel states also play prominent roles in the experimental behavior.
Despite several decades of theoretical effort, it remains a major
challenge to achieve a fundamental and predictive understanding
within a microscopic framework of the thermodynamics, structure,
and dynamic properties of the homogeneous fluid phase, and the
equilibrium and nonequilibrium phase transitions, over the wide
parameter range of experimental relevance.

A primary focus of this feature article is the structure and
scattering patterns of athermal colloid-polymer mixtures. This is
one of the most poorly understood aspects of these systems, and
little information is available from experiment, theory or
computer simulation. The only theoretical approach presently available
that can treat structural correlations of all species at the molecular
level is liquid state integral equation methods. Thus, we shall
concentrate on our recent contributions \cite{16,17} in    this area based
on the polymer reference interaction site model (PRISM) approach
\cite{18,19,20,21,22}. We first summarize the current level of    theoretical
understanding to place our work in proper context.        

The earliest
theoretical model of athermal polymer-colloid    suspensions was
developed by Asakura and Oosawa (AO) \cite{23}. They    addressed the
most elementary question of the effective    entropy-driven "depletion
attraction" between two hard spheres    dissolved in a polymer
solution.  Several major simplifications    were introduced: (i)
$R \gg R_{g}$ , (ii) polymer coils were treated    as hard spheres in
their interactions with the large colloids    thereby ignoring
internal conformational degrees of freedom, and    (iii)
polymer-polymer interactions and correlations were ignored    and
hence the polymer solution was effectively an ideal gas. Point
(iii) is appropriate when $c\ll c^{*}$ , or $c<c^{*}$ if the
polymers are in special $\Theta$ solvents where the    polymer-polymer
second virial coefficient vanishes. The physical    origin of the
attraction can be viewed in two distinct, but    equivalent,
manners. When colloidal surfaces approach to less than    a separation
$\sim 2 R_{g}$, the polymer is "squeezed out" of the    gap between
the particles resulting in an unbalanced polymer    osmotic pressure
exerted on the colloids thereby pushing them
together. Alternatively, the polymer (treated as a sphere with no
internal structure) can increase its (translational) entropy if    the
particles cluster and share the excluded volume depletion    layer
surrounding each of them.        

For the above restrictive AO
conditions, the finite colloid    concentration problem can be
approximately treated by adopting an    effective 1-component fluid
model where the polymers enter    implicitly via an (assumed) pair
decomposable AO depletion    potential. If one accepts the AO
simplifications (i)-(iii) then    3-body and higher effective
interactions can be ignored if $R_{g}/R    \le 0.154$. Considerable
progress in understanding the competition    between fluid-fluid and
fluid-crystal phase transitions was made    using this approach by
Gast, Hall and Russel \cite{24}. However, as    recently emphasized by
Dijkstra, Evans and coworkers \cite{25},    besides the obvious
limitations, effective 1-component models    cannot be rigorously
derived from the true binary mixture with    {\em interacting}
polymers since no small parameter exists {\em    even} in the classic
$R_{g} \ll R$ regime.        

In the opposite limit to the AO model,
when the colloidal    particles can be considered as small point
disturbances in the    much larger polymer coils, integrating out the
polymer degrees of    freedom in the presence of few particles is
tractable \cite{26} 
and has lead    to deep insights \cite{27,28,29}. This is possible
because such point    disturbances can be treated perturbatively; for
random walk coils the
change in free    energy by one particle of size $R$ is of order $R
R_g^2 c \ll 1$.    Whether this knowledge will lead to a theory for
the particle    structure remains open because polymer induced
many-body    interactions among the nanoparticles appear crucial.

If the polymers are treated as literally noninteracting ideal
random walk coils, and polymer concentration is assumed to be
vanishing small, then a tractable 2-component statistical
thermodynamic theory can be constructed within the simplified AO
model framework \cite{30}. It is formally equivalent to the
Widom-Rowlinson  nonadditive hard sphere mixture problem \cite{31}, and
the "phantom sphere free volume" theory of Lekkekkeker et. al.    \cite{30}
has had additional successes in qualitatively predicting    polymer
partitioning and 3-phase equilibria in certain system    parameter
regimes. However, this approach has been recently shown
experimentally to incur significant quantitative and qualitative
errors for thermodynamic properties and fluid-fluid phase diagrams
even when $R_{g}<R$ \cite{32,33}. Moreover, it is inappropriate for
higher polymer concentrations $c > c^{*}$, and/or smaller    particles
where $R_{g} > R$.        

Indeed, it is not clear what one means by
"depletion" if $R_{g}    \gg R$ and polymers can "wrap around"
particles (see Fig. 1).    Besides the obvious fact that pair
decomposable effective    colloidal potentials lose meaning,
interpretation in terms of    "squeezing polymer out" between surfaces
and unbalanced osmotic    pressure felt by colloids seems
inapplicable. However, the    fundamental idea that the perturbation
of polymer coils by hard    particles can be reduced, and polymer
conformational entropy    enhanced, by the physical clustering of
particles remains valid.    But, proper theoretical description now
demands confronting the    internal conformational degrees of freedom
of the polymer coils,    and how entropy loss due to excluded volume
interactions with    colloids depends on particle-particle separation.

\begin{figure}[ht]    \centerline{{\epsfysize=7.cm
\epsffile{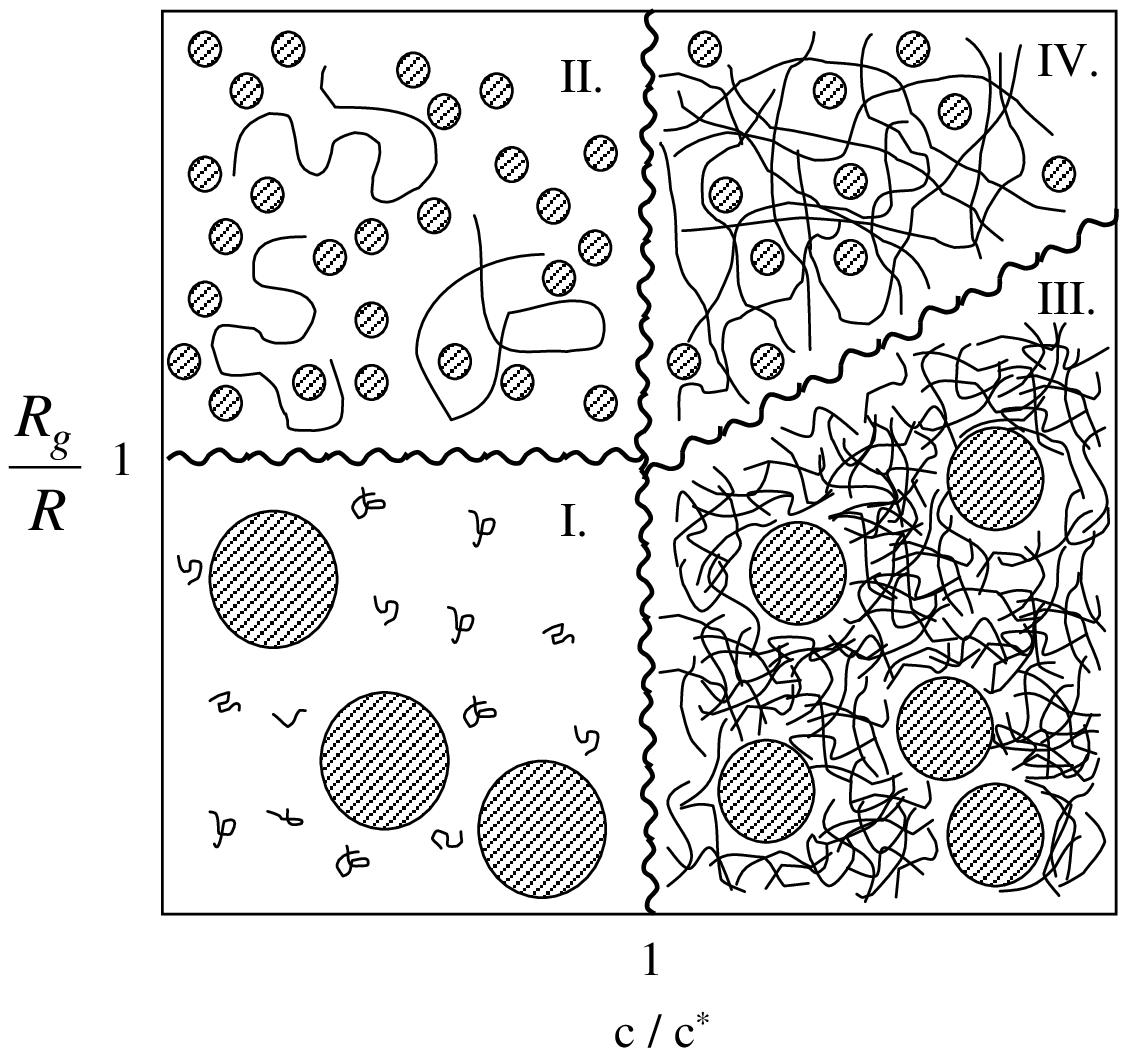}}} 
\caption{Schematic illustration of the 4
distinct physical    regimes. Dilute and semidilute polymer solution
regimes are    divided according to      $R_g \ll R$ (I) or $R_g \gg
R$ (II), and $\xi \ll R$ (III) or    $\xi \gg R$ (IV), respectively
\label{fig1}. }    \end{figure}

Thus, we believe an essential challenge is to construct a
microscopic 2-component theory which can address within a common
framework the 4 different physical regimes shown schematically in
Fig. \ref{fig1}. This requires treating polymers realistically as
connected chains of units (monomers or segments) which experience
excluded volume forces among themselves, and with particles, at    the
elementary segmental scale. The role of polymer internal
(conformational) degrees of freedom, or monomer density
fluctuations, would then be directly taken into account, and its
consequences for nonideal polymer solution behavior, including
physical mesh formation at $c > c^{*}$, would emerge naturally.
Moreover, no pair decomposable depletion potential approximation
would be necessary, particle penetration of the polymer coil would
be accounted for, and segment-segment and segment-colloid spatial
correlations could be predicted for the first time. A general goal
of our recent work has been the formulation of such a theoretical
approach which can give a unified description of thermodynamic
properties, phase behavior, and structure. Knowledge of the latter
is of intrinsic interest, and also provides essential input into
modern microscopic theories of the dynamics of such suspensions    and
their  gelation or vitrification \cite{34,35,36,37,38}.      

  To achieve the above
goals requires combining statistical    mechanical ideas from the
fields of polymer and colloid physics.    The great range of relevant
length scales, and geometric    asymmetries between the species, have
rendered computer simulation    of polymer-colloid  mixtures at this
level nearly impossible    \cite{39,40,41}. An emerging attempt at a
2-component description which    approximates  polymer chains as "soft
colloids" is interesting    \cite{42,43}. However, such an approach avoids
the explicit treatment    of polymer internal degrees of freedom, is
not applicable if    $R_{g} > R$, and the subtle question of what
effective    polymer-colloid pair potential to employ has not yet been
addressed. We have chosen to pursue a microscopic liquid state
theory approach based on atomic and macromolecular integral
equation methods, in particular the PRISM theory \cite{19} which is the
polymeric generalization of the RISM theory of Chandler and
Andersen \cite{20,21} developed for small, rigid molecular fluids. The
connection between polymer integral equation theory and
(coarse-grained) Hamiltonian based Gaussian field theory with
constraints has been established \cite{22}. The PRISM approach has been
extensively developed and widely applied over the past decade \cite{19}
including the successful treatment of flexible and semiflexible
polymer solutions, melts, alloys, self-assembling block    copolymers,
liquid crystalline polymers \cite{44}, and star-branched    macromolecules
\cite{45}.       

 There are at least two major challenges to formulating a
reliable,    computationally convenient, liquid state theory of
athermal    polymer-colloid suspensions. First, a fully numerical
based    approach which works at the segment level encounters severe
difficulties spanning the 3-4 orders of magnitude of length scales
(if $R \gg R_{g}$) from segment to colloid diameter \cite{46}.
Moreover, local chemical detail on the segmental scale may not be
important for mesoscopic particles and polymer concentrations well
below the semidilute-concentrated crossover $c^{**}$. The latter
concentration signifies the point at which the universal    structural
feature of a semidilute polymer solution, the physical    mesh or
polymer density-density screening length, becomes    comparable to, or
smaller than, the monomer size \cite{6}. For $c\ll    c^{**}$, which is the
regime relevant to most suspension    experiments, a field theory
inspired and analytically tractable    "thread model" of flexible
polymer chains has been previously    developed and is employed here
\cite{19,47,48}.     

   The second major challenge is of a conceptual
nature. It is well    known from atomic liquid state theory that
reliable prediction of    structure, and especially phase separation,
of highly asymmetric    (but additive) hard sphere mixtures is a
difficult problem due to    its sensitivity to closure approximations
\cite{34}. In particular, the    classic Percus-Yevick (PY) theory
incorrectly predicts the absence    of fluid-fluid phase separation
\cite{49}. In mixture problems,    accurate treatment of the "cross
correlations" is a very important    and demanding task. For the
present problem, this involves the    packing of polymer segments (not
entire coils) against mesoscopic    particles. We have found that to
properly capture the essential    physics within a liquid state theory
framework requires a new    approximation scheme (called "modified
PY", m-PY) that accounts    for nonlocal entropic repulsions between
polymer segments and the    colloid (due to polymer connectivity
correlations and    conformational perturbations) in a predictive,
thermodynamically    self-consistent manner.       

 The goal of this
article is not to present the technical details    of the m-PY PRISM
theory which are available elsewhere \cite{16,17}.    Rather we first
summarize the essential theoretical ideas in    section 2, and then
concentrate on the structural consequences for    which the liquid
state approach makes unique predictions. Prior    theoretical results
of others for structural questions are limited    to the elementary,
but still physically complex, problem of 1 or 2    particles dissolved
in a polymer solution. The polymer    segment-colloid density profiles
for a single hard sphere in a    dilute or semidilute polymer solution
have been exhaustively    studied for all $R_{g}/R$ size ratios
\cite{27,28,29}. Field theory,    scaling arguments, density functional theory
and self-consistent    mean field theory have all been utilized to
varying degrees and    for restricted size asymmetry regimes to study
the problem of    polymer-induced depletion interactions between a
pair of particles    at infinite dilution \cite{27,50}. These results serve
as valuable    benchmarks to test the reliability of the PRISM
approach in the    special limiting situations as discussed in section
3. The    opposite dilute regime, 1 polymer in a particle suspension,
is    also considered in section 3, and is not treatable by scaling or
self-consistent mean field approaches.        We then turn to the
question of polymer-colloid mixtures with both    components present
at nonzero concentrations. Our predictions    \cite{16,32} for spinodal
fluid-fluid phase separation are summarized    in section 4 and
contrasted with prior theories. In sections 5-7,    we present new
representative results for all three structural    pair correlation
functions, on both local and mesoscopic length    scales, and in both
real and Fourier space. Their physical    interpretation is
emphasized, and the systematic dependence on    controllable system
variables is established. The paper concludes    in section 8 with a
discussion of the present limitations of the    theory, and an outlook
to the future.        Finally, we shall make only brief contact with
prior theoretical    studies. Detailed comparisons for structural
correlations are not    possible since the vast majority of prior work
is of a statistical    thermodynamic nature, and polymer correlations
at nonzero    concentrations have not been addressed.

\section{PRISM - mPY  theory}      

  We consider the density
fluctuations as quantified by the partial    structure factors
$\hat{S}_{ij}(q)\sim \langle \delta\varrho_i^*({\bf    q})
\delta\varrho_j({\bf q})\rangle$  for wavevector $\bf q$, of a
binary mixture of polymer chains (composed of $N$ interaction    sites
or segments) and hard spheres (radius $R$) which interact    solely
via pair decomposable site-site hard core repulsions    between all
species. In obvious matrix notation, the total density    fluctuations
in Fourier space consist of single molecule    contributions described
by a (diagonal) intramolecular structure    factor, $\hat{\omega}_{ij}(q) =
\hat{\omega}_{j}(q) * \delta_{ij}$, and    intermolecular site-site
correlations, $\hat{h}_{ij}(q)$, 
\beq{eq1}    \hat{S}(q) =
\hat{\omega}(q) + \varrho^{1/2}\;  \hat{h}(q)\;    \varrho^{1/2}\; . \eeq 
The diagonal matrix of densities,    $\varrho_{ij} =
\varrho_{j} \delta_{ij}$, defines the number    density of colloids
and polymer segments.  The intermolecular pair    correlation
functions, $g_{ij}(r) = 1 + h_{ij}(r)$, describe the    relative
probabilities of finding a site of species $i$ at a    distance $r$
from a site of species $j$ at the origin. The    intermolecular hard
core exclusion constraints require $g_{ij}(r<
(\sigma_{i}+\sigma_{j})/2) = 0$, where $\sigma_{2} = \sigma_{c}$    is
the colloid hard core diameter, $\phi_{c} = (\pi/6) \varrho_{c}
\sigma_{c}^{3}$ the associated packing fraction, and $\sigma_{1} =
\sigma_{p}$ is the excluded volume diameter of a single polymer
site (segment); polymer coils, which can interpenetrate,
nevertheless cannot intersect their backbones. Note that
intramolecular excluded volume enters via the single polymer chain
structure factor.        

The generalized Ornstein-Zernike, or
Chandler-Andersen, equation    is given by \cite{20,21}: 
\beq{eq2}
\hat{S}^{-1}(q)  =    \hat{\omega}^{-1} -  \varrho^{1/2} \, \hat{c}(q)
\,    \varrho^{1/2}\; . \eeq 
which serves to define the effective
potentials, the site-site direct correlation functions,
$c_{ij}(r)$. Liquid state theory approaches are based on the
classic idea that the $c_{ij}(r)$ are relatively simple objects
that exhibit a spatial range which is of the order of the    potential
range even if the system builds up collective long    ranged
fluctuations. The familiar random phase approximation (RPA)    \cite{6}
replaces the $c_{ij}(r)$ with the bare pair potentials, but    fails
to fulfill the excluded volume constraints on $g_{ij}(r)$.      

  Use
of these exact formulae is made in PRISM theory by invoking    several
approximations which we view as physical assumptions that    can be
tested, either directly by experiments or simulations, or    via
comparisons of their consequences with more rigorous    alternative
theories in tractable limiting cases. PRISM is    fundamentally a
theory for structural properties deduced not from    an approximated
free energy, but from approximated equations for    the structure
factors and pair correlation functions themselves.    Besides the well
established use of such liquid state approaches    \cite{19,21,34}, and the
absence of alternative methods with comparable    reaches, our
discussion of the m-PY PRISM approach is further    motivated by its
prediction of a scaling law limit for large    polymers, which is an
extension of known field theoretic results    for dilute systems. The
connection of the corresponding structural    correlations to this
aspect and to the physical approximations    will be one of our
central topics. The model and statistical    mechanical approximations
used in m-PY PRISM are the following:    
\begin{itemize}
\item[$(i)$]    The rigid colloid is described by a single site,
leading to    $\hat{\omega}_{c}(q)$ =1. The structure of a single flexible
polymer is    assumed to be the same everywhere in the system and to
be known a    priori (``homogeneous pre-averaging assumption'' \cite{21,22}
).    
\beq{eq3} \hat{\omega}_{p} \equiv \omega(q) = \frac 1N
\sum_{\alpha\beta}^N \; \langle e^{i \vek{q} ( \vek{r}_\alpha -
\vek{r}_\beta ) } \rangle \approx \frac{N}{ ( 1 + q^2 \xi_0^2)} \;
, \eeq 
The second approximate equality in \gl{eq3} corresponds to
adopting an analytically simplified ideal (Gaussian) chain
description where the coil radius-of-gyration $R_{g} = \sqrt 2
\xi_{0}$, or equivalently $R_{g} = \sqrt{2N} l_p$, where $l_p$ is
proportional to the ideal walk step, or "statistical segment"
length.  Additionally, only site averaged quantities are
considered, and    hence specific chain-end effects are ignored \cite{19}.
\item[$(ii)$]    For the colloidal hard sphere and athermal polymer
components, we    adopt the accurate site-site PY approximation:
$c_{jj}(r>\sigma_{j}) = 0$, stating that the effective potentials
vanish beyond overlap of the sites. This strict implementation of
the original Ornstein-Zernike idea has shown its great value for
the description of the pure systems \cite{19,21,34}, and improvements    on
it are available.    
\item[$(iii)$]    For the polymer segment-colloid
direct correlation,  we have    proposed the novel m-PY approximation
as a one parameter extension    of the PY closure \cite{16}: 
\beq{eq4}
\hat{c}_{cp}(q)=    \frac{\hat{c}^s_{cp}(q)}{1+q^2\lambda^2}\;,
\quad\mbox{with}\quad    c^s_{cp}(r>\frac{\sigma_c+\sigma_p}{2})=0\;,
\eeq or equivalently    in real space \beq{eq5} c_{cp}(\vek{r}) =
\int\!\!d^3s\;    \frac{1}{4\pi\lambda^2} \; \frac{1}{
|\vek{r}-\vek{s}|}\;    e^{-|\vek{r}-\vek{s}|/\lambda}\;
c^s_{cp}(\vek{s})\; . \eeq 
The    function $c_{pc}^s(r)$ is the analog
of the short range PY direct    correlation function which vanishes
for segment-colloid separation    beyond contact, and can be
interpreted as describing unconnected    polymer segments. It must be
determined such that the exact core    condition for $g_{cp}(r)$ is
satisfied. The parameter $\lambda$    introduces the spatial
nonlocality of the segment-colloid    interactions due to entropic
considerations. The spatial    convolution form in \gl{eq5} describes
nonlocal conformational    constraints on segmental packing (chain
connectivity) within a    distance $\lambda$ of the colloidal
surface. Such a medium-ranged    effective interaction becomes
necessary in theories employing a    pre-averaging assumption for the
single polymer structure since a    polymer coil rearranges close to a
repulsive surface. Neglect of    these rearrangements (equivalent to
the standard $\lambda = 0$    site-site PY approximation) within PRISM
results in a qualitative    overestimate of the polymer segment
density close to the    particles, thereby entailing a severe
underestimate of the induced    depletion attraction \cite{17,52}.
\end{itemize} 
   The m-PY PRISM approach adopts further technical
approximations or    simplifications when handling specific properties
of the system.    
\begin{itemize}    
\item[$(iv)$]    Polymer
correlations at low semidilute concentrations obey    universal
scaling laws, which are known from scaling arguments and    field
theoretic calculations \cite{6}. PRISM exhibits an equivalent    limit,
termed the "thread" model \cite{47,48}, which is analytically    tractable
and can be considered an approximation to the scaling    laws,
physically appropriate for dilute and semidilute polymer
concentrations, but quantitatively slightly different. The polymer
specific scaling laws express that excluded volume interactions
between segments on different coils remain active even if the
thickness of the segments is negligible, $\sigma_{p}\to 0$. Only
the polymer segment density relative to the semidilute crossover
segmental number density (which defines when polymer coils just
begin to interpenetrate and interact), $\varrho_p^{*}$, is then
relevant for long polymer chains. The dimensionless measure of
polymer concentration is $\varrho_{p}/\varrho_{p}^{*} = c/c^{*} =
(4 \pi/3) (\varrho_{p}/N) R_{g}^{3}$.    
\item[$(v)$]    Our
additional assumption of Gaussian single chain statistics in
\gl{eq3} corresponds to a mean-field approximation to the polymer
correlations with self-avoidance and entails the familiar errors    in
the scaling law exponents as thoroughly discussed elsewhere
\cite{6}. This approximation is fundamentally different to the often
used literal ``ideal chain'' approximation, which also neglects
polymer-polymer interactions. While the former simplifies polymer
correlations, the latter totally neglects polymer interactions
treating the coils as an ideal gas. As discussed later, this
generally fails for higher polymer densities and/or $R_g$.
\item[$(vi)$]    The information required to solve the 3 coupled PRISM
integral    equations is now specified given the nonlocality
parameter,    $\lambda$. Significant analytic progress can be made
using the    Baxter or Wiener Hopf factorization method \cite{17,34}. The
parameter    $\lambda$ contains much of the many body physics of
depletion    interactions and is expected to vary nontrivially with
all system    parameters, $\phi_{c}$, $R_{g}/R$ and $c/c^{*}$. In
order to    achieve a parameter free theory, a thermodynamic
consistency    condition is enforced. The excess (nonideal) chemical
potential    for inserting a single polymer ($\varrho_{p}\to 0$) into
a    colloidal suspension of volume fraction $\phi_{c}$ is required to
agree when computed from the compressibility and free energy
charging routes. 
 The two formally exact expressions for this
quantity are equated yielding a highly nonlinear equation for
$\lambda$ [21]: 
\begin{eqnarray} \label{eq6sss} - \int_0^{\varrho_c}
d\varrho'_c\; N\; \hat{c}_{cp}(q=0,\varrho_c')|_{\varrho_p=0}       &
= &    \frac{\pi \varrho_c\sigma_c N}{2}\; \int_0^1\!\!\!\! d\zeta
(\sigma_p\!\!+\!\!\zeta\sigma_c)^2 g^{\rm
(\zeta)}_{cp}(\frac{\sigma_p\!\!+\!\!\zeta\sigma_c}{2})|_{\varrho_p=0}
\nonumber\\ & + &    2 \pi \varrho_c^2\sigma_c^3 N\; \int_0^1\!\!\!\!
d\zeta \zeta^2    \frac{\partial g^{\rm
(\zeta)}_{cc}(\zeta\sigma_c)}{\partial    \varrho_p}|_{\varrho_p=0}\;
.    \end{eqnarray}  
  The compressibility route emphasizes long
wavelength information    encoded in the segment-colloid direct
correlation function via    charging up (thermodynamic path
integration) of the colloid    density. On the other hand, the free
energy route emphasizes the    local contact value of the real space
segment-colloid pair    correlation function via a growth process
whence the diameter of    the colloidal spheres is increased from zero
to its full value.    Requiring $\lambda$ be chosen so that these two
very different    exact routes yield identical polymer insertion
chemical potentials    is a powerful consistency constraint on the
structural correlation    functions.        Exact numerical
determination of $\lambda$ from the above    equations is difficult
because thermodynamic integrations are    required. Thus, based on
exact analytic analysis of several    limiting cases, we proposed an
approximate interpolation formula    which is quite accurate
(satisfies thermodynamic consistency to    within 15\% or less for all
parameter values) \cite{16,17}: 
\beq{eq7}    \lambda^{-1}= \xi_0^{-1} +
\frac{1+2\phi_c}{1-\phi_c}\frac{\lambda_1}{\sigma_c}\; , \eeq  
  where
$\lambda_{1}= 1 + \sqrt5$. The limit $\phi_{c}\to 0$    corresponds to
the (still many body) problem of one polymer and    one colloid. In
nearly quantitative agreement with mean field    theoretic studies
\cite{27,28,29,50}, and physical intuition, the    nonlocality length scale is
then predicted to reduce to the    polymer correlation length
($\xi_0=R_{g}/\sqrt2$) if  $R\gg    R_{g}$, and a fraction of the
colloid radius, $2R/\lambda_1$, if    $R_{g} \gg  R$. For $R_{g}\ll
R$, $\lambda$ is nearly independent    of colloid concentration since
the coils can fit into the voids    between particles and screening of
the depletion layer around one    colloid by other particles is not
effective. Such a situation is    implicit in using  one-component
colloid models with effective    ($\phi_{c}$-independent) depletion
attraction pair potentials    between particles. However, for
$R_{g}\gg R$, long polymers can    wrap around the colloids, and
$\lambda$ is predicted to    monotonically decrease with increasing
colloid volume fraction.    This reflects many particle screening of
the repulsive    polymer-colloid interactions by colloidal density
fluctuations,    which reduces entropic attraction resulting in a
filling in of the    depletion layer surrounding the colloidal
particles.    
\item[$(vii)$]    To determine $\lambda$ beyond
vanishing small polymer    concentrations, we have employed the "blob
scaling" concept well    known in polymer physics \cite{6}. The idea is
that when $c>c^{*}$ in    semidilute solution the physical mesh, or
equivalently polymer    density-density correlation or screening
length, $\xi$, is the    relevant length scale, not $R_{g}$. Note,
that $\xi=R_g\,    \hat{\xi}(R_g/R,c/c^*,\phi_c)$ is explicitly
calculated within our    theory and not obtained from blob scaling. In
the absence of    particles, thread PRISM theory predicts \cite{19}:
\beq{eq8} \frac 1\xi    = \frac{1}{\xi_0}+ 4\pi\, \rho_p l_p^2=
\frac{1+\frac{3}{\sqrt8}\,    \frac{c}{c^*}}{\xi_0}\; , \eeq 
If one
employs the known scaling    law for the effective statistical segment
length in semidilute    athermal solutions, $l_p\sim\rho_{p}^{-1/8}$,
then \gl{eq8} agrees with    experiment, field theory, and scaling
arguments for athermal    polymer solutions \cite{6,19}. In the limit of
$\phi_{c}\to 0$, at    nonzero polymer concentrations, $\xi_0=R_{g}/\sqrt2$
in \gl{eq7} is    replaced by the mesh length, i.e. $\xi_{0}\to \xi$. For
$R\gg    R_{g}$, the depletion layer and nonlocality parameter are
predicted to be given by $\xi$, in agreement with field theory and
scaling arguments for a polymer solution near a large colloid or
effectively a planar surface \cite{28,50}. At nonzero colloid
concentration, the effective polymer correlation length, $\xi$,
will depend on $\phi_{c}$. A naive idea would be that since the
colloids occupy finite fraction, $\phi_{c}$, of the available
space, the semidilute crossover concentration, $c^{*}$, would
decrease by the factor of $(1-\phi_{c})$ in \gl{eq8}. Numerical
studies \cite{16,17} of the full m-PY equations reveals this idea is
very accurate in the colloid regime of $R_{g}\ll R$. However, in
the opposite extreme nanoparticle regime, this naive idea does not
hold\footnote{We find $\frac{\xi_0}{\xi}       =
1+\frac{3}{\sqrt8}\frac{c}{c^*}
\frac{1-0.42\phi_c}{(1-\phi_c)(1+2\phi_c)}$ for $R_g\gg R$, which  
gives an initial increase  of $c^*(\phi_c)$ by    23\%.},
and \gl{eq8} remains quite accurate       implying an insensitivity of
$\xi$ to colloid volume fraction for very long polymers.
\end{itemize}     

   We now turn to the systematic examination of
structural    correlations. Common system parameter choices are made
to allow    cross comparisons between the three distinct correlation
functions    and partial scattering structure factors. We shall
present results    for three values of $R_{g}/R = 0.1$ (representative
of common    colloid systems), 1.0 (the crossover case), and 5 (easily
achievable with surfactant micelles, dendrimers or proteins as the
"nanoparticle"). Results for a "moderate" and "high" value of
colloid packing fraction are discussed, $\phi_{c}=$ 0.2 and 0.45,
respectively. Analytic results in the asymptotic $R_g/R    \rightarrow
\infty$ and $R_g/R \rightarrow 0$ limits are also    presented.

Finally, we point out that comparison of experimental scattering
patterns with our predictions is direct, and no convolutions with    a
single polymer form factor are necessary since we model polymers    at
the elementary segmental scale; for a discussion of an alternative
approach via polymer-center-of-mass correlations see
\cite{krakoviack}. Here, for example, the total    scattering
function is given by the standard formula: \beq{eq9}    I(q) =
\varrho_c b_c^2 P(q) \; S_{cc}(q) + \varrho_p\, b_p^2 \;    S_{pp}(q)
+ 2 \sqrt{\varrho_c\varrho_p} \, \sqrt{P(q)}\, b_c b_p    \;
S_{cp}(q)\; , \eeq where $b_c$ and $b_p$ are the colloidal and
polymer segment scattering lengths, respectively, and $P(q)$ is    the
particle form factor.  The traditional emphasis is on    extracting
the colloidal partial structure factor, $S_{cc}(q)$. We    hope our
calculations will motivate future neutron scattering    experiments to
directly measure the polymer scattering functions.

\section{Dilute  limits}  

  The dilute limits are relatively well
understood and provide    important insight into the structural
modifications in finite    composition mixtures. We focus on the
structural aspects, and    refer the reader elsewhere \cite{17} for the
thermodynamic results.      

  \subsection{Dilute particles}     

\begin{figure}[ht]
\centerline{{\epsfysize=7.cm        \epsffile{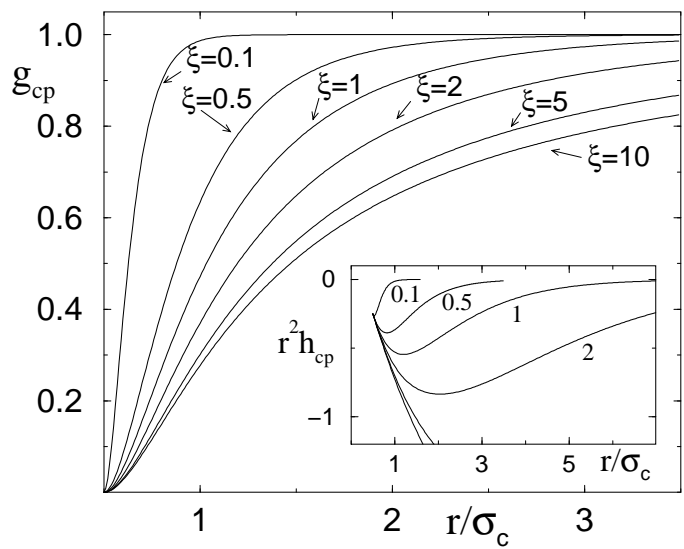}}}
\caption{Polymer-colloid pair correlation function, $g_{cp}(r)$,    in
the dilute and semidilute polymer concentration region    exhibiting
the  polymer segment depletion layer close to an    isolated
colloidal sphere for various polymer correlation    lengths $\xi$ (in
units of $\sigma_c$) as labeled. The inset shows    the same data as
$r^2 h_{cp}(r)=r^2 (g_{cp}(r)-1)$ which reveals the long ranged    tail of the
depletion layer for large $R_g/R$. \label{fig2} }    \end{figure}

   In
this limit the polymer correlations are unaffected by the    presence
of particles. The polymer solution is characterized by a
density-density correlation (or screening) length, $\xi$, also
known as the physical mesh length or blob size. In dilute polymer
solutions, $\xi$ reduces to $R_{g}/\sqrt 3$ ($R_{g}/\sqrt2$ within
our simple approximation of \gl{eq3}). The polymer segment density
at a distance $r$ from a particle is given by $\varrho_p\,
g_{cp}(r)$. Representative results for the polymer segment-colloid
radial distribution function are given in Fig. \ref{fig2}. All display    the
classic "depletion hole" suppression below the random value of
unity and a monotonic approach to unity as $r \to\infty$. There    are
three important aspects of these results.     

\noindent   (1) Close to contact,
the pair correlations grow in a power-law    fashion, $g_{cp}(r)\sim
(\frac{r-R}{\lambda})^{1/\nu}$ (with    Flory-exponent $\nu=\frac 12$ in
our mean-field like description),    as required from the wall-Virial
theorem \cite{27,28,29,50}. It connects    the osmotic pressure of the polymer
solution ($\Pi\sim    \frac{kT}{\xi^3}$ for semidilute concentrations)
to the force per    area that the segments exert. The force is simply
proportional to    segment density, which is $\varrho_p\,
g_{cp}(R+\sigma_p)\sim    \varrho_p\, (\sigma_p/\xi)^x$. Equating the
pressures determines    $x=1/\nu$ \cite{28,50}. Gratifyingly, the various
theoretical estimates    of the (universal) coefficients of the
quadratic growth law agree    to within $\sim20\%$  error in both the
$R \gg R_{g}$ extreme    colloid and $R_{g} \gg R$ extreme
nanoparticle limits \cite{17}.\\
(2) In the extreme colloid regime,
the particle appears to the    polymer as a flat wall. The width of
the depletion layer is the    polymer correlation length, $\xi$.\\
(3) In the extreme nanoparticle limit, the polymer-particle    density
profile changes qualitatively. Two length scales emerge. A    local
length of order the particle radius due to the direct    perturbation
of (local) polymer conformational entropy by the    particle, and a
longer length scale of order $\xi$. For    intermediate separations,
the profile increases in a $\sim r^{1/\nu-d}$    power law manner
reflecting the self-similar chain connectivity    correlations on
length scales smaller than the mesh size. The    inset in Fig. \ref{fig2}
highlights the new longer length scale feature    which clearly
emerges when $\xi \gg  R$. Whereas the existence of    a narrow (of
order R) component of the depletion layer for $R_g\gg    R$ is rather
obvious, the aspect that a long polymer (considering    the dilute
case for simplicity) cannot totally balance the    disturbance of a
point repulsion on distances shorter than $R_g$    warrants further
discussion, especially since its consequences appear
 to have    been overlooked in
several prior works \cite{53,54,55}. It is equivalent    to the statement that
of order $N$ segments of the chain {\em    rearrange} when a small
particle is added. The free energy    increase when adding a few
particles is proportional to the number    of segments 
{\em independently displaced} 
by each of the particles. It    can be estimated from the average
number of polymer segments within the sphere volume, which needs to be
corrected by the number of correlated segments belonging to the same
polymer strand.
Thus it scales like $\delta F/(V    kT) \sim \varrho_p
\varrho_c \sigma_c^{d-1/\nu} l_p^{1/\nu} \sim    c_p \varrho_c \, (N
\sigma_c^{d-1/\nu} l_p^{1/\nu})$ where $l_p$    is the Kuhn-segment
length and the latter expression reveals the    dependence on the
polymer molecule number density \cite{29,54,sear}. Therefore,    on a per coil
basis the free energy increase is of order $N$ which    requires the
whole chain to rearrange  around the particle. The    cross second
virial coefficient in this limit,    $B_2^{cp}=\partial^2 \left(\delta
F/(V kT)\right)/(\partial c_p    \partial \varrho_c) = -    \frac12
\hat{h}_{cp}(q=0)\sim N \sigma_c^{d-1/\nu} l_p^{1/\nu}$    thus grows
with $N$, and as it is given by the integral of the    correlated part
of the segment profile, $g_{pc}(r)$ can approach    unity only for $r
\sim R_g$.      

  The extreme nanoparticle regime results were
previously known only    from the field theoretic work of Eisenriegler
and coworkers    \cite{27,28,29}. Moreover, their experimental significance
was only very    recently realized from the studies by Kulkarni et al.
\cite{56} of the    second virial coefficient of small proteins dissolved
in polymer    solutions. These experiments discovered a novel
nonmonotonic    dependence of the consequences of depletion
attractions on polymer    concentration which originates in the
different concentration    dependences of the two depletion layer
lengths. Sensitivity to    solvent quality, and hence polymer-polymer
correlation length,    were also discovered. The experimental
observations have been    shown to be well described by PRISM theory
\cite{56}.    

    It is significant to note that alternative polymer-based
treatments of the nanoparticle regime have come to conflicting,    and
apparently incorrect, conclusions. The early analysis of de    Gennes
\cite{54} suggests the depletion attraction effect was    negligible, and
all mixtures were miscible, if $R < \xi$, thereby    seemingly missing
the long range aspect discussed above. Recent    work by Odijk \cite{55},
and Tuinier et al. \cite{53} come to similar    conclusions that 
the range of depletion    attractions between
two spheres is controlled solely by particle    size and is $\sim3R$.
These workers employ an uncontrolled    "superposition of 1-particle
depletion layer approximation", which    is not invoked in the field
theoretic or integral equation    approaches. The particle-particle
second virial coefficient for    ideal coils is reported to remain
positive under dilute    ($c<c^{*}$) polymer conditions \cite{53}, in
disagreement with the    field theoretic \cite{27} and PRISM \cite{17,56}
predictions, and the    experimental observation \cite{11,13,15,32} of
fluid-fluid demixing at    $c<c^{*}$ when $R_{g} > R$.  

\begin{figure}[ht]    
\centerline{{\epsfysize=7.cm
\epsffile{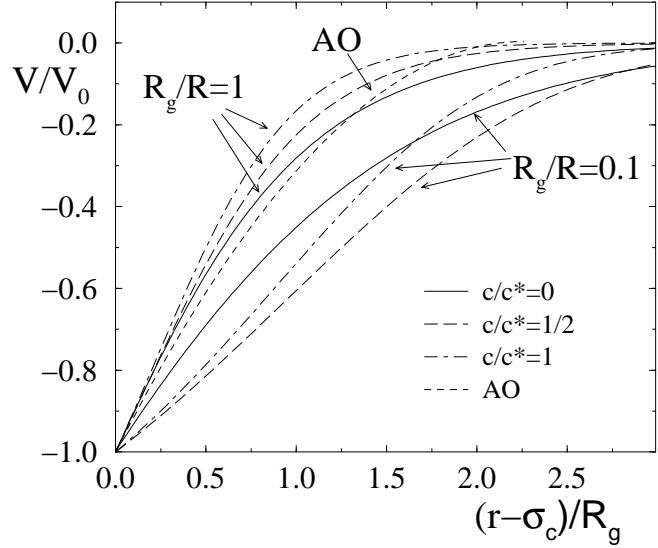}}}   
 \caption{Normalized polymer induced
potential of mean force,    $V(r)/V_0=-        \log{(g_{cc}(r))}/V_0$,
where $V_0=\log{g_{cc}(\sigma_c)}$, versus reduced    distance,
$(r-\sigma_c)/R_g$, in the limit of vanishing colloid    concentration
and for the labeled polymer concentrations. Thick    lines correspond
to $R_g/R=0.1$ and thin lines to $R_g/R=1$. The    short dashed line
shows the Asakura-Oosawa result with a depletion    layer of
$\frac{2}{\sqrt\pi} R_g$. \label{fig3} }    \end{figure}

      The
classic depletion effect is generally discussed at the level    of an
effective entropic attraction, $V(r)$, between a pair of    particles
mediated by the polymer solution. If $c \ll c^{*}$, it    is the
theoretical foundation for effective 1-component and    related
statistical thermodynamic approaches. Within the PY    closure for
particle-particle direct correlations, m-PY PRISM    theory yields, in
units of $kT$, a result showing clearly the    connection to the
polymer correlations: 
\beq{eq10sss} V_{cc}(r)=- \ln    ( 1 + W(r)
),\qquad W(r) = \Theta(r-2R)    \int\!\!\!\frac{d^3q}{(2\pi)^3}
e^{-i{\bf qr}} \varrho_p    c_{cp}^2(q) S_{pp}(q) \; .\eeq
$W(r)=g_{cc}(r)-1$ is analytically    available but sufficiently
cumbersome that we refrain from    presenting it here. Alternatively,
if one adopts the hypernetted    chain (HNC) closure for
colloid-colloid direct correlations, one    obtains $V(r) = - W(r)$
\cite{17,52}. In the $c/c^{*} \to 0$  ultra    dilute polymer limit, the
HNC and PY based expressions agree.    Hence, we analyze this regime
to make contact with the classic AO    model. The latter is formulated
in the $R\gg R_{g}$ regime, for    which one obtains: 
\beq{eq11}
V_{AO} = \frac 32\;    \frac{c}{c^{*}}\; \frac{R}{R_{g}}\;  [1 - H
]^{2} \; , \eeq 
where    $H = ( r - 2R ) / 2 R_{g}$ is the
surface-to-surface separation of    the large colloids, and the
spatial range of $2R_{g}$ follows    directly from the modelling of a
polymer coil by a hard sphere of    radius $R_{g}$. Based on
self-consistent mean field results \cite{50},    it is sometimes suggested
that the width of the depletion layer    within an AO framework should
be $\sqrt{4/\pi} R_{g}$. This small    quantitative modification
increases $V_{AO} (2R)$ by a factor of    1.27, and increases the
spatial range of the depletion potential    by 12\%. The m-PY PRISM
result in the analogous limit is \cite{17}:   
 \beq{eq12} V_{\rm mPY} =
\frac{27}{8}\; \frac{c}{c^{*}}\;    \frac{R}{R_{g}}\;  [ 1+ ( 5/9 ) X
+ ( X/3 )^{2} ]\; e^{-X} \; ,    \eeq 
where  $X = ( r - 2R ) /
\xi_{0}$. The PRISM and AO depletion    potentials are qualitatively
identical in the $R\gg R_{g}$ ultra    dilute polymer
limit. Quantitatively, the amplitude of the PRISM    result is roughly
a factor of 2 larger than AO, and its range is    modestly longer (see
Fig. \ref{fig3}). Both these differences are    connected to the fact that
PRISM  treats the polymer as a    fluctuating coil and
(approximatively) accounts for  the loss of    both translational and
conformational (orientational)  entropy    when confined between two
large colloids.      Field theory also includes the latter physical
effect which results   in a nonzero V(H) beyond H=1 and a contact
strength which is larger   than the AO result \cite{28}:
$V(H=0)=3(c/c^{*})(R/R_g)\ln{2}$.          

  Figure \ref{fig3} presents results
for the normalized potential of mean    force, $-kT \ln{g_{cc}(r)}$, of
m-PY PRISM at several reduced    polymer concentrations and two
$R_{g}/R$ values. The $R_{g}/R=0.1$    case is quite close to the
extreme colloid limit discussed above.    Interestingly, the spatial
range of the PRISM depletion potentials    is a nonmonotonic function
of reduced polymer concentration,    reflecting a subtle and
nontrivial dependence of the amplitude at    contact and polymer
correlation length on $c/c^{*}$. Only for    larger polymer, (e.g.,
$R_g=R$ in Fig. \ref{fig3}) does the expected    trend emerge that the
potential range decreases as $\xi$, the blob    size.        

 From a
broad theoretical perspective, we believe the qualitative    agreement
between the field theoretic and m-PY PRISM approaches    (where as far
only mean-field exponents could be analytically    handled) for the
polymer-particle density profiles (and    thermodynamics \cite{16,17}) over
the entire range of $R_{g}/R$    provides a solid foundation for
extension of the integral equation    theory to the nonzero $c/c^{*}$
and $\phi_{c}$ mixture regimes of    primary experimental interest.

\subsection{Dilute polymers}    

    The opposite problem of one or two
polymers in a hard sphere fluid    is far less understood \cite{6,26}. The
central statistical geometry    problem is how polymer coils can "fit
into" the free volume of a    colloidal fluid, which in the dilute
polymer limit is not    perturbed by their presence. When $R_{g} > R$,
the popular    approximation of an open, fractal polymer coil by a
solid particle    is clearly invalid.     

\begin{figure}[ht]    
\centerline{{\epsfysize=7.cm
\epsffile{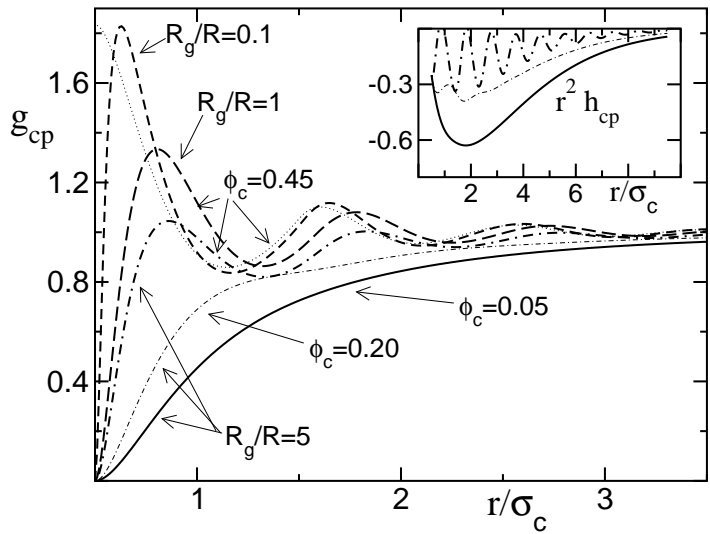}}}  
  \caption{Polymer-colloid pair correlation
functions, $g_{cp}(r)$,    for dilute polymers ($\varrho_p=0$) and
colloid packing fraction    and size ratio as labeled. The (unlabeled)
thin dotted line gives    the asymptotic result for $R_g/R\to0$
corresponding to Eq.    (\protect\ref{eq12s}). The inset shows the
long-ranged part of the    depletion layer for the size ratio
$R_g/R=5$ (line style as in the    main panel) which is apparent in a
plot of $r^2\, h_{cp}(r)=r^2\,    (g_{cp}(r)-1)$. \label{fig4} }
\end{figure}

   Examples of the
polymer-colloid profiles are given in Fig. \ref{fig4}. As    expected, at low
colloid volume fraction the behavior is similar    to that shown in
Fig. \ref{fig2}. For large polymers, the depletion hole    has a local and
long range component. As the colloids are    densified, the local
depletion layer narrows due to many particle    screening of the
direct perturbation of the polymers by the    repulsive
polymer-colloid interactions. The dependence of the    depletion layer
on colloid density renders simple effective    potential approaches
inapplicable in general. In the limit of    vanishingly small
polymers, $R_g\ll R$, the polymer correlations    can be explicitly
found. The segment density jumps from 0 to    $\varrho_p/(1-\phi_c)$
at contact. Thus for  $R_g/R\to0$ the    polymer correlations directly
follow the colloidal ones,    
\beq{eq12s} \hat{h}_{cp}(q) =
\frac{-1}{1-\phi_c}\; \hat{F}(q) \;    \hat{S}^{\rm HS}(q)\wer
\hat{F}(q)=\int\!\!d^3r e^{i{\bf qr}}\,    \Theta(R-r)\;. \eeq 
$\hat{F}(q)$
is the scattering amplitude of the empty    polymer regions whose
centers are correlated according to  the    hard sphere structure
factor $\hat{S}^{\rm HS}(q)$ of the    particles. The particle form
factor is given by    $P(q)=|\hat{F}(q)/(4\pi R^3/3)|^2$, and also
appears in the dilute    polymer correlations,
$\hat{S}_{pp}(q)=\frac{4\pi}{3}R^3 \varrho_p
\frac{\phi_c}{(1-\phi_c)^2}\; P(q)\; \hat{S}^{\rm HS}(q)$ for
$R_g\ll R$.   Hence, at high colloid volume fraction there appears
an oscillatory packing of the polymers on the colloid size length
scale. This represents an "imprinting" of the colloidal structure
and free volume correlations on the polymer segment spatial
organization. Considering the polymers as phantom spheres, this effect
was first observed by Louis et al. \cite{42}.
 As Fig. 4 shows, where the limit corresponding to
\gl{eq12s} is included, it is present for all $R_{g}/R$ ratios    even
in the nanoparticle regime where the polymers must "go    around" many
particles in the fluid. However, careful examination    of $g_{cp}(r)$
curves for $R_g\gg R$ (see the inset) reveals the    long range
component of the depletion hole is not  destroyed, but    only becomes
of smaller relative amplitude and more difficult to    resolve.

\begin{figure}[ht]    
\centerline{{\epsfysize=7.cm
\epsffile{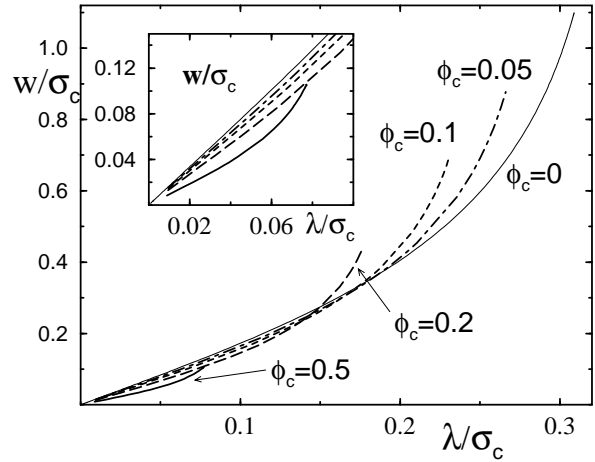}}}  
  \caption{ Local width of the
polymer segment depletion layer    defined by
$g_{cp}(\frac{\sigma_c}{2}+w)=\frac12$ versus the    colloid-polymer
interaction length $\lambda$ for the labeled    colloid concentrations
and at vanishing polymer concentration.    Along the curves, the
polymer size varies between $0.03\le R_g/R    \le 280$, and $\lambda$
is given by Eq. (\protect\ref{eq7}). The    result for vanishing
colloid concentration, thin solid line, also    describes the width at
finite polymer concentration, where the    polymer correlation length
is given by Eq. (\protect\ref{eq8}).    The inset presents an enlarged
view relevant for small polymer    correlation lengths. \label{fig5} }
\end{figure}

A summary of our attempt to quantify the local width of the
depletion layer is given in Fig. \ref{fig5}. The partial collapse of the
results over a 4 orders of magnitude variation of size asymmetry
suggests a strong connection between the layer width, $w$, and the
fundamental nonlocality parameter, $\lambda$. The relatively    modest
variation of the width is remarkable. At higher colloid    volume
fractions the layer width becomes a unique function of    $\lambda$,
which only mildly splays out as $\phi_{c}$ is    decreased. Both $w$
and $\lambda$ saturate at finite values in the    limit $R_{g}/R \to
\infty$ since the relevant length scale is then    the colloid
size. Polymer-polymer correlations in the dilute limit    are
discussed in section 7.        

\section{Fluid-fluid phase separation}

\begin{figure}[ht]    
\centerline{{\epsfysize=8.cm
\epsffile{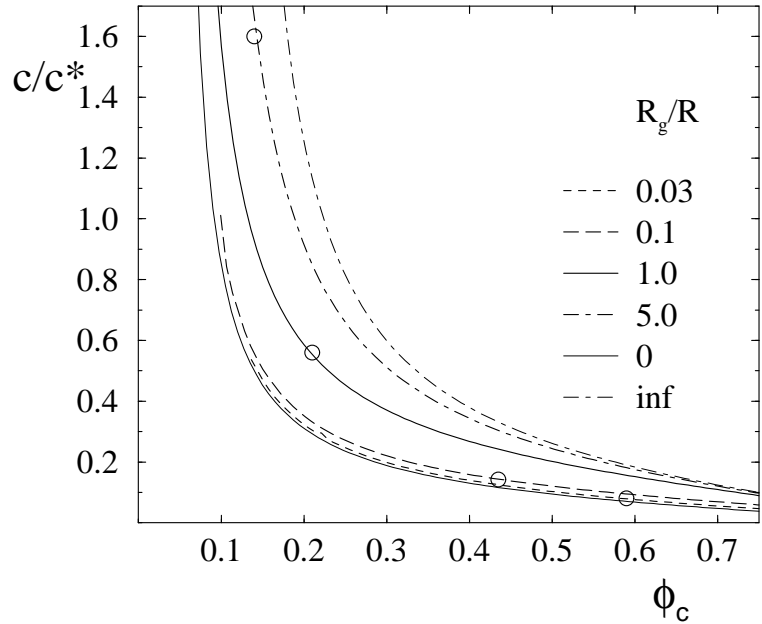}}}   
 \caption{Spinodal curves for various size
ratios. The critical    points are marked by circles, and thin lines
indicate the  limits    for small (thin solid) and large (thin
dot-dashed) polymer to    colloid size ratios $R_g/R \to 0
(\infty)$, respectively.    \label{fig6} }    \end{figure}

The fluid-fluid spinodal instability lines, and corresponding
critical points, predicted by m-PY PRISM theory are shown in    Fig.
\ref{fig6}. The novel trend is that the homogeneous mixed phase is
stabilized as the ratio $R_{g}/R$ increases. As discussed in great
depth elsewhere \cite{16,32}, the predicted trend is in agreement with
recent experiments on model athermal polymer-colloid suspensions
when fluid-fluid phase separation is the thermodynamically stable
transition ($R_{g}/R > 0.3$). However, all the classic theories    we
are aware of \cite{23,24,30}, and the recent density functional    approach
\cite{58}, predict the opposite shift of fluid-fluid phase    boundaries
with size asymmetry. The use of the PRISM dilute    polymer insertion
chemical potential in the classic theories does    not correct this
qualitative error \cite{32}. Thus, the essential    importance of nonzero
polymer concentration, for which there are    direct and (colloid)
induced polymer-polymer interactions, has    been unequivocally
established \cite{32}.    

    In the extreme colloid limit, $R \gg  R_{g}$,
the spinodal curves    approach a limiting form. This might be
interpreted as suggesting    an effective 1-component, simple liquid
type approach is adequate    for small polymers, perhaps of the very
short range adhesive hard    sphere (AHS) type \cite{59}. However, the
critical point of the AHS    model is located at $\phi_{c}
\approx0.1$, versus the prediction    of 2-component PRISM of a very
concentrated value. Indeed, PRISM    theory predicts the critical
point shifts to higher colloid    concentrations as $R/R_{g}$
increases, in qualitative agreement    with the phantom sphere free
volume model \cite{30}, the behavior of    highly asymmetric hard sphere
mixtures \cite{25}, and 1-component    attractive square well or Yukawa
fluid models \cite{59,60}.       

 Another generic feature of Fig. \ref{fig6} is
the absence of spinodal    phase separation at sufficiently small
colloid volume fraction of    $\phi_{c} = 1/22 \approx 0.045$. The
origin of this prediction is    that with decreasing $\phi_{c}$ the
value of reduced polymer    concentration required for phase
separation exceeds $c^{*}$. Since    we have considered athermal, good
solvent conditions for the    polymers, when $c > c^{*}$ repulsive
interchain interactions    systematically reduce the spatial range of
correlated polymer    density fluctuations \cite{6}, and hence depletion
effects, resulting    in the predicted behavior. For $R \gg  R_{g}$
and low $\phi_{c}$    the depletion potential between two particles is
generally large    compared with the thermal energy. Hence, the issue
of closure    approximation for the colloid-colloid direct
correlations (HNC    versus PY) becomes relevant \cite{17}. Such a concern
also applies to    all prior classic free energy approaches
\cite{23,24,30}, including    recent density functional theories
\cite{58}.

The spinodal predictions of PRISM theory for the extreme
nanoparticle regime, $R_{g} \gg  R$, also contain several novel
aspects. The qualitative failure of the AO model is not surprising
\cite{24}, and alternative polymer-based approaches are relatively few.
The de Gennes and Tuinier et al. \cite{53,54} approaches discussed in
section 3.1 ignore the long range component of the polymer mediated
attraction. This results in a massive weakening of the depletion
effect in the nanoparticle regime, and suggests no phase    separation
occurs if  $R \ll  \xi$ and $c < c^{*}$. However, this    disagrees
with Fig. \ref{fig6}
 (and experiments \cite{11,13,14,32} ) which    predicts phase
separation is possible when $R \ll  \xi$ and $c <    c^{*}$.

Sear has recently presented a simple theory in the phantom sphere
free volume model spirit for the $R \ll  R_{g}$ case \cite{61}. Polymer
coils were taken to be noninteracting and ideal, even well into    the
semidilute regime \cite{6} where demixing is predicted with a    $\phi_{c}$
at the critical point that decreases rapidly to zero    proportional
to the ratio $(R/R_{g})^{2}$ due to "many body    effects". This
latter prediction contradicts Fig. \ref{fig6} where the    critical colloid
volume fraction deep in the semidilute regime    approaches a limiting
value of 0.12, typical of a van der Waals    type fluid experiencing
weak, long range attractions. We believe    the qualitative
discrepancy is due to the Sear theory not taking    into account
direct polymer-polymer repulsions and physical mesh    formation for
$c > c^{*}$ , and colloid-induced polymer-polymer    attractions at
finite polymer concentrations. The direct polymer    repulsions are
most important in athermal good solvents and when    $R_{g} \gg  R$,
but are also  present even for so-called    $\Theta$ solvent
conditions where ideal polymer solution behavior    fails dramatically
above $c^{*}$ \cite{6}.     

   Finally, in the extreme nanoparticle limit
the spinodal curves in    Fig. \ref{fig6} also approach a limiting
value. This behavior arises from    the prediction of m-PY PRISM
theory that the colloids induce an    attractive second virial
coefficient between dilute polymers which    scales with coil volume
\cite{17}. Physically, the extreme nanoparticle    limit is akin to the
classic polymer-solvent problem \cite{6}, where    the critical point is
well known to be correlated with $c^{*}$.    For our present problem,
the "solvent" is a mesoscopic particle    and the effective
colloid-induced attraction between polymer    segments is now
$\phi_{c}$ dependent. Since the extreme    nanoparticle regime is
characterized by long range, but weak,    depletion effects we believe
our prediction is very reliable.   

  We now return to our major
focus, structural correlations at    finite compositions of both
components. Polymer concentration will    generally be reported in
dimensionless units reduced by the value    at the spinodal,
$c_{s}$. Figure \ref{fig6}
 provides the required    information to convert this
to the classic reduced variable,    $c/c^{*}$. Results for a value of
$c/c_{s}$ close to unity will be    emphasized in order to establish
the maximum realizable structural    changes in the homogeneous phase.

\section{Colloid-colloid correlations}

\begin{figure}[ht]    \centerline{{\epsfysize=8.cm
\epsffile{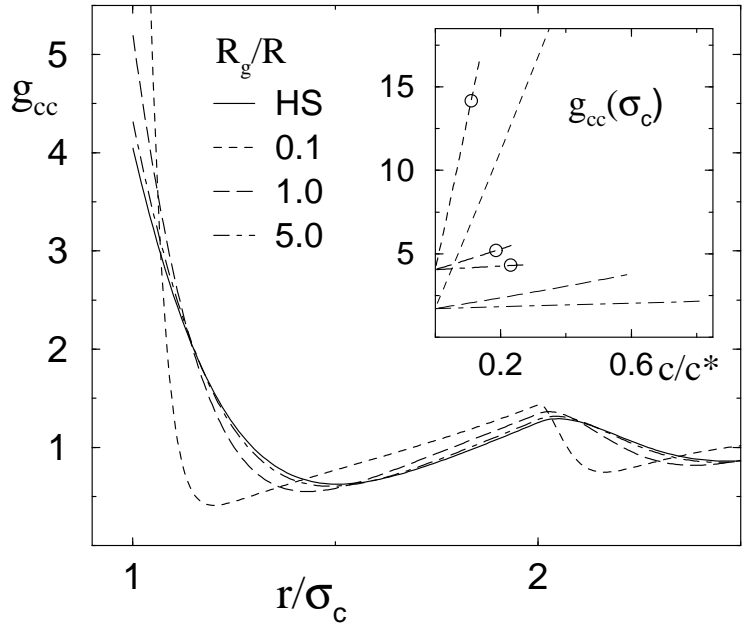}}}    \caption{Colloid pair correlation
functions, $g_{cc}(r)$, at    colloid packing fraction $\phi_c=0.45$
and polymer concentration    at a fixed relative distance to the
spinodal, $c/c_s=0.8$, for the    size ratios as labeled; the solid
line (HS) gives the result for    hard spheres. The thick lines in the
inset give the contact values    for the three polymer sizes as
function of the polymer    concentration; circles mark the ones of the
main figure. Thin    lines indicate the corresponding contact values
at $\phi_c=0.20$.    \label{fig7} }    \end{figure}

        An example of the
influence of polymers on colloid real space    packing is given in
Fig. \ref{fig7} for several values of size asymmetry.    The case shown is a
high colloid volume fraction, but the basic    trends are the same at
lower values of $\phi_{c}$. The primary    effect is the expected
enhanced local clustering of hard spheres    due to the
polymer-mediated depletion attraction. This phenomenon    is enhanced
(decreased) in amplitude (spatial range) as $R_{g}/R$    decreases
consistent with classic ideas about the role of    attractions on
fluid structure due to van der Waals and their    modern formulation
by Weeks, Chandler and Andersen (WCA) \cite{62}. For    the smallest
polymer case corresponding to a very short range    attraction, the
form of $g_{cc}(r)$ begins to resemble the classic    AHS model result
\cite{63}, including the highly distorted second    solvation shell
feature. In the large polymer limit, one can show    that the
deviation between $g_{cc}(r)$ and its pure hard sphere    analog is
proportional to  $R/R_{g} \ll  1$. This makes    visualization of
structural changes in the nanoparticle regime    difficult, but the
long range depletion mechanism discussed in    section 3 is still
present and observable in wavevector space (see    below). Explicitly,
we find for $R_g/R\to\infty$ 
\beq{eq13}    g_{cc}(r) \to g^{\rm
HS}(r/\sigma_c) + b\; \frac{R}{r}\;    A(\phi_c)^2\; e^{-r/\xi_c},
\qquad b=4 \pi R \,    l_{p}^{2}\rho_p=\frac{3}{2}\, \frac{c}{c^{*}}
\, \frac{R}{R_g}    \eeq 
where $g^{\rm HS}$ is the PY hard sphere pair
correlation    function, and the (positive) Lorentzian term describes
the long    ranged attraction induced structure. It gets suppressed
with    increasing $R_g$, and colloid packing fraction since
$A(\phi_c)=(1+\frac{2}{\lambda_1})\frac{1-\phi_c}{1+2\phi_c}$.    Only
the leading terms on small or large scales, $r/\sigma_c$ or
$r/\xi_0$, respectively, are given. The collective correlation
length, $\xi_c$, increases from the polymer correlation length to
infinity when approaching the mean-field spinodals: 
\beq{eq14}
\frac{\xi_0}{\xi_c} = 1 - \frac{3}{\sqrt8}\, \frac{c}{c^*}\;
(f(\phi_c)-1)\qquad \mbox{where }\quad    f(\phi_c) = \frac{\phi_c(6
\lambda_1 + 1 - 4    \phi_c)}{2(1-\phi_c)(1+2\phi_c)}\; . \eeq 
The
quantity $f(\phi_c)$    is the segment-segment contact value for
segments from two    different coils, which increases from its value 0
without    particles, to large values with increasing $\phi_c$; see
Fig. 15    in section 7. These $R_g \gg R$ structural results again
suggest    the polymer induced depletion can be considered in a van
der Waals    like perturbative spirit, and that its amplitude is small
and    given by the locally replaced polymer segment density,
$\varrho_p    l_p^2 R$. This parameter also determined the chemical
potential    for inserting a single sphere into a polymer fluid as
discussed in    section 3.1. The intuitive picture, supported by 
m-PY (and    field theory results for $\phi_c=0$) is that this
dimensionless    interaction parameter depends linearly on polymer
concentration    throughout the dilute and semidilute concentration
region since it    arises from the interaction of the particle with an
isolated    polymer strand. It gets renormalized only when the polymer
mesh    size becomes comparable to the particle size, $\xi \cong R$.
Hence, we are rather confident that \gl{eq13} describes the    colloid
structure qualitatively up to values of $c/c^*$ such that    $\xi
\cong R$.      

  The enhancement of the contact value of $g_{cc}(r)$
increases    monotonically and essentially linearly with polymer
concentration,    as seen in the inset of Fig. \ref{fig7} for two colloid
volume fraction    cases. It is interesting to observe that the
enhancements relative    to the pure hard sphere value are larger for
the lower $\phi_{c}$    case. This is a consequence of the increased
ability of short    range attractions to induce local reorganization
when the fluid is    more compressible. The linear dependence of the
contact value on    polymer concentration is presumably related to our
use of the PY    closure for the colloid-colloid direct correlations,
and hence may    be less reliable when $R \gg  R_{g}$ in analogy with
prior studies    of hard sphere mixtures \cite{49,57}. 

\begin{figure}[ht]    \centerline{{\epsfysize=8.cm
\epsffile{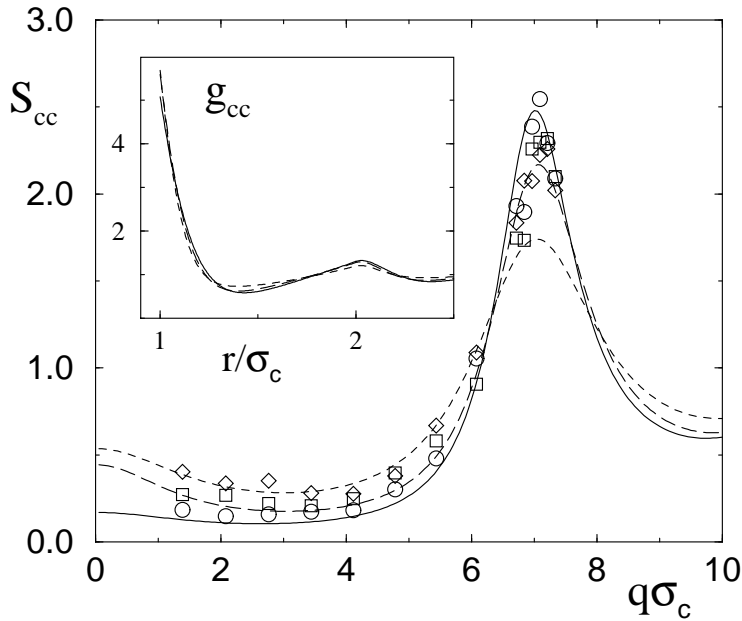}}}    \caption{ Dimensionless colloid structure
factors compared to  the    experimental data of Moussaid    et
al. \protect\cite{7}.    The parameters are
$(\phi_c,c/c^*,R_g/R)=$ (0.333, 0.13, 0.24;    short dashes and
$\diamond$), (0.404, 0.13, 0.37; long dashes and    ${\scriptstyle
[]}$), and (0.444, 0.10, 0.57; solid line and    $\circ$). The inset
shows the corresponding colloid pair    correlation functions
$g_{cc}(r)$ with the same line styles.    \label{fig8} }
\end{figure}

The colloidal
structure factor    displays the classic wide angle peak at $q_{p}
\approx 7    \sigma_{c}^{-1}$ which is a quantitative measure of the
degree of    coherent  order of the local "cage". A zero adjustable
parameter    comparison of PRISM theory with recent light scattering
measurements \cite{7} on model athermal polymer-colloid suspensions
under near triple point conditions is given in Fig. \ref{fig8}. Good
agreement is found at both small and large wavevectors.
Earlier analysis of one-component pair-potential theories required 
fits with strongly underestimated polymer concentrations as pointed
out in \cite{42}, where a polymer-phantom-sphere approach achieved 
descriptions of comparable quality to Fig. \ref{fig8} upon adjusting
$c/c*$ somewhat. The
deviations around the wide angle peak for the smallest polymer
sample require further investigation due to possible experimental
difficulties. 
 The inset shows the corresponding predictions for    the
colloidal radial distribution function. Interestingly, despite    the
fact that the three samples are all at different values of
$\phi_{c}$, $c/c^{*}$, and $R_{g}/R$, the $g_{cc}(r)'s$ nearly
collapse onto a single curve. This may be a consequence of the
experiments being performed near the triple point of each system.
Future video microscopy experiments can perhaps test this
prediction.       

 As phase separation is approached, very large
enhancements of the    long wavelength fluctuations emerge, which can
be quantified by    the dimensionless colloidal osmotic
compressibility, $S_{cc}(0)$.    The agreement between theory and
experiment in Fig. \ref{fig8} at small    wavevectors is very significant
since $S_{cc}(0)$ is nearly an    order of magnitude larger in the
polymer-colloid suspensions than    the analogous pure hard sphere
fluid \cite{7}. Very recently,    quantitative comparisons of osmotic
compressibility data obtained    using the turbidity technique have
been obtained by Ramakrishnan    et al. \cite{33} for a large set of model
athermal polymer-colloid    suspensions. Significant errors of both a
quantitative and    qualitative nature are incurred by the classic
statistical    thermodynamic approaches, while PRISM theory provides a
quantitative or semiquantitative, no adjustable parameter
description of the measurements as $R_{g}/R$ was varied from 0.026
to 1.4 \cite{33}.     

\begin{figure}[ht]    \centerline{{\epsfysize=8.cm
\epsffile{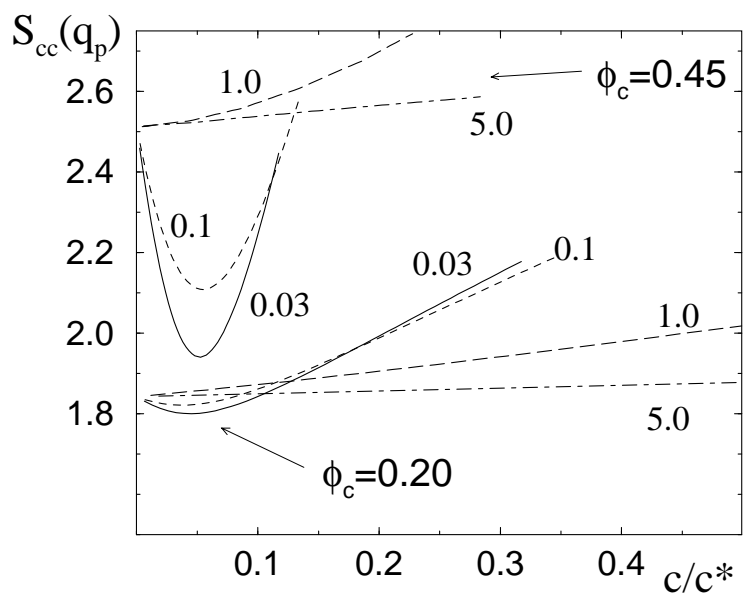}}}    \caption{Large-angle scattering peak
intensity, $S_{cc}(q_p)$,    versus polymer concentration for two
packing fractions,    $\phi_c=0.45$ and 0.20, and for the labeled
polymer-colloid size    ratios, $R_g/R$. The curves at $\phi_c=0.20$
are shifted upwards    by $0.6$. \label{fig9} }    \end{figure}

   The wide angle scattering peak intensity
$S_{cc}(q_{p})$ plays the    role of a local collective order
parameter, and displays rich and    subtle trends as a function of
polymer concentration and mixture    size asymmetry (see Fig. \ref{fig9}).
There are at least two distinct    physical processes. At high colloid
packing fraction, the local    order is initially significantly
depressed upon polymer addition    if the coils are small compared to
the particle. In this regime    the polymers are readily dissolvable
in the colloidal fluid, and    the short range depletion attraction
makes the colloids "sticky"    which decreases the coherence of the
local cage making it more    heterogeneous. However, this trend
dramatically reverses at high    polymer concentrations as phase
separation is approached. Here the    classic depletion effect
corresponding to local clustering and    densification of the colloids
comes into play as a precursor to    phase separation, resulting in
enhancement of the local order    parameter. With increasing relative
polymer size, the initial    "cage melting" process becomes
inefficient due to the longer range    of the effective depletion
attraction, and only the latter process    is present. In the
nanoparticle limit, very little structural    modification is
observed, consistent with van der Waals and WCA    ideas. A similar
sequence of  behaviors is observed at the lower    colloid packing
fraction in Fig. \ref{fig9}, but the "cage melting"    process is greatly
reduced as expected.     

   Changes of the local $S_{cc}(q)$ and
$g_{cc}(r)$ upon polymer    additions can have major dynamical
consequences, which in the    colloidal regime far from a glass
transition can be estimated from    various statistical dynamical
approaches of the generalized Enskog    type \cite{38}.  With increasing
colloid volume fraction, a    well-studied hard sphere glass
transition occurs in the absence of    polymers at
$\phi_c\approx0.57$. The microscopic idealized mode    coupling theory
(MCT) has been successfully applied to understand    this phenomenon
\cite{35}. For pure hard spheres, the essential idea is    that a density
is reached where the local cage constraints,    quantified by the wide
angle $S_{cc}(q)$, become sufficiently    strong that particles become
localized on a Lindemann length scale    of ca. 0.1 $\sigma_{c}$. This
is predicted to occur at a $\phi_{c}    \approx 0.516$ and
$S_{cc}(q_{p}) \approx 3.54$. A simple, but    reliable, approximate
implementation of the ideal MCT is to use    the latter condition on
the local order parameter as the defining    criterion for glass
formation. If $R_{g}/R \ll  1$, then an    effective 1-component model
to treat the dynamics seems sensible    \cite{35,36,37}. Hence, the influence
of polymer additives on the value of    $\phi_{c}$ at the colloidal
glass transition can be predicted    using the PRISM structural
information.    

\begin{figure}[ht]    \centerline{{\epsfysize=8.cm
\epsffile{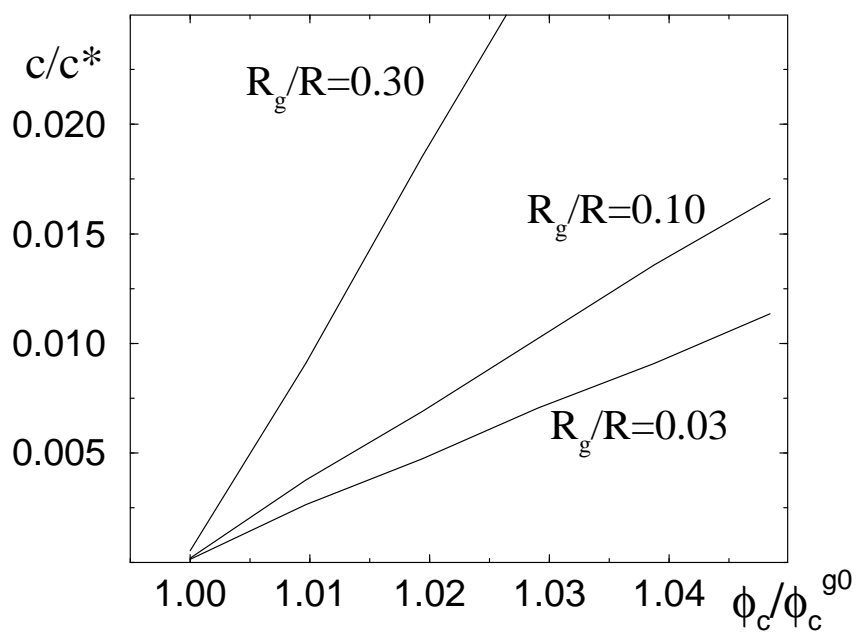}}}    \caption{Shift of the glass transition
line, relative to the value    for hard spheres,
$\phi_c^{g0}=\phi_c^g(c/c^*=0)=0.516$ (theory)    or
$\phi_c^{g0}=0.58$ (experiment) as estimated from
$S_{cc}(q_p)=3.54$ versus polymer concentration for size ratios as
labeled. \label{fig10} }    \end{figure}

    Sample results are shown in Fig. \ref{fig10}. One sees
that remarkably    small concentrations of polymer can at high density
significantly    reduce local cage ordering, resulting in major
increases of the    glass transition volume fraction which nearly
follow a linear    dependence with $c/c^{*}$. A shift of up to 10\% is
experimentally    relevant before the random close packing constraint
is    encountered. The polymers can effectively "melt" a glass, and
this    phenomenon has been experimentally observed in model hard
sphere    colloid-polymer mixtures \cite{8}. As $R_{g}/R$ increases, the
ability    of polymers to disrupt the local hard sphere packing is
reduced,    and the glass melting mechanism is less effective. For
$R_{g} = R$    (not shown), there is essentially no perturbation of
the glass    volume fraction.  With increasing $c/c^{*}$, the lines in
Fig. \ref{fig10} "turn around" well before the fluid-fluid spinodal
boundary is    encountered, and the glass suppression effect is
reversed. This    curious nonmonotonic trend is expected from the
shape of    $S_{cc}(q_{p})$ at high colloid density shown in Fig.
\ref{fig9}. We do    not show this behavior here, since it is now established
within    MCT that at increasing attraction strength a new
non-ergodicity    transition emerges that has a gelation type
character with    particles localized on the length scale of the
attractive    interaction ($R_{g}$) and not $\sigma_{c}$ \cite{36,37}. Such
a system    is sometimes called a "attraction-driven" glass in
contrast to the    classic jamming effect in hard sphere colloids
driven by purely    repulsive packing considerations, and cannot be
predicted based on    the simple $S_{cc}(q_{p}) = 3.54$ criterion
valid for (nearly)    hard sphere fluids and weakly attractive
colloids \cite{36,37}.        

\section{Polymer-colloid correlations}

\begin{figure}[ht]
\centerline{{\epsfysize=7.cm        \epsffile{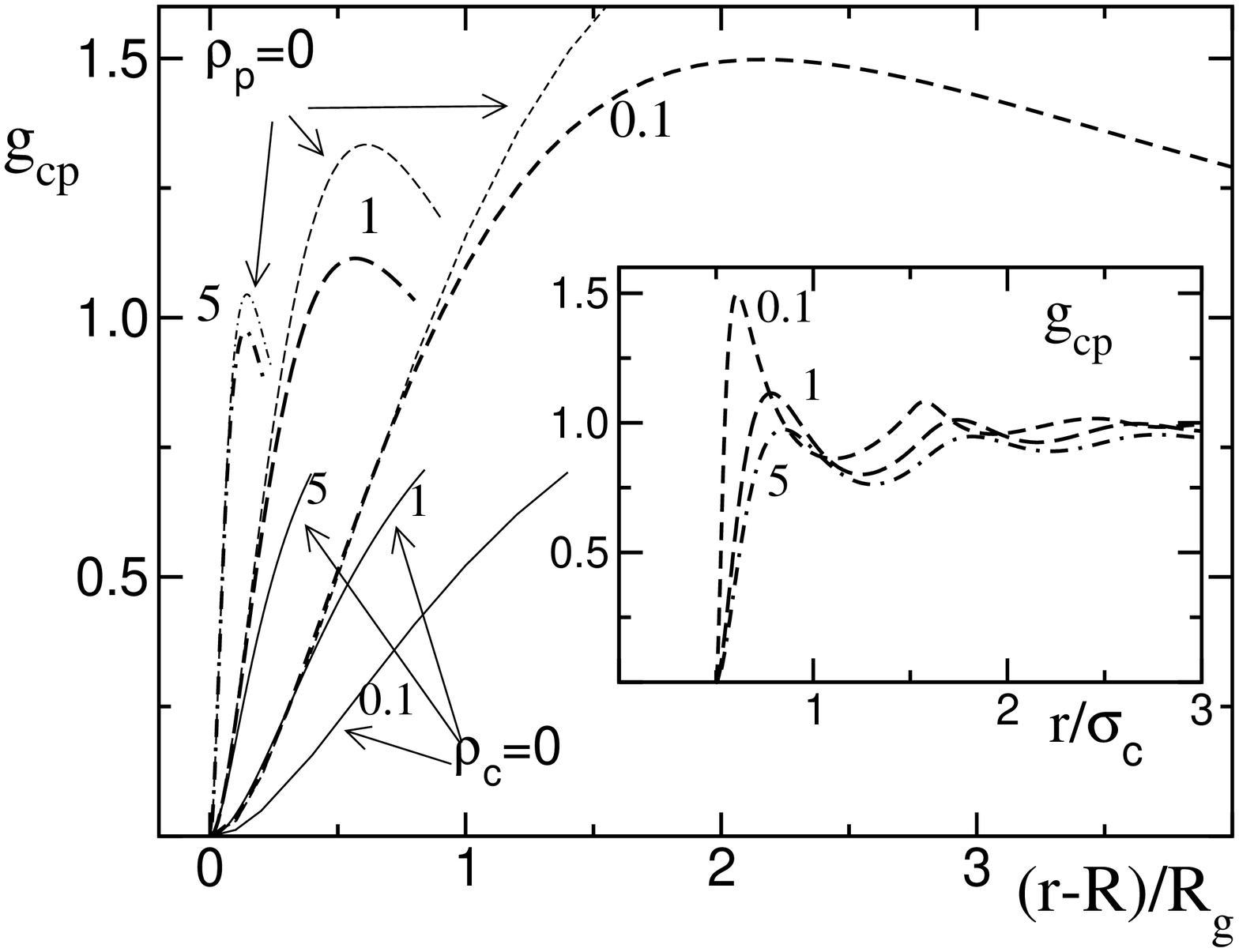}}}
\caption{ Polymer segment density profile, $g_{cp}$, versus
rescaled distance, $(r-R)/R_g$, from a particle surface; size
ratios, $R_g/R=0.1$ (short dashes), 1.0 (long dashes) and 5.0
(dot-dashed) as labeled. As in Fig. \protect\ref{fig7} the colloid
concentrations $\phi_c=0.45$, and the polymer concentration
relative to the spinodal is constant, $c/c_s=0.8$. Thin lines of
the same styles give the results for the same $R_g/R$ and $\phi_c$
but at vanishing polymer concentrations, $c/c_s=0$. Thin solid
lines show the depletion layer for identical polymer
concentrations,  $c/c_s=0.8$, but at vanishing colloid
concentration, $\phi_c\to0$,  at the studied size ratios as
labeled. The inset shows the finite density curves versus
$r/\sigma_c$ out to larger distances. \label{fig11} }    \end{figure}

In this section we investigate the influence of simultaneous
nonzero polymer and colloid concentrations on the polymer
segment-colloid pair correlations. Figure \ref{fig11} presents
representative results for a high colloid volume fraction and    three
values of $R_{g}/R$. The limiting cases of zero polymer
concentration, and zero colloid volume fraction, are shown for
comparison. The most striking result is how weakly dependent the
packing correlations are on polymer concentration, especially in
the more local depletion hole regime where segments are within a
distance $R_{g}$ of the particle surface. This arises because the
polymer coils are rather compressible at the considered densities
below the semidilute threshold concentration, $c^{*}$. The peak in
$g_{pc}(r)$ on macromolecular length scales ( $r-R > R_{g}$ ) is
suppressed as polymer concentration increases, and more so as the
colloid size increases. A physical interpretation of this trend is
that as $c/c^{*}$ increases the individual identity of spherical
polymeric coils is reduced due to (weak) interpenetration of
different chains. The situation may be akin to polyelectrolyte
solutions which at low salt concentration pack in a simple liquid
like manner which gradually disappears as polymer concentration
and/or Debye-Huckel screening length decrease \cite{64}. However, the
physics of inhomogeneity suppression in the neutral    polymer-colloid
mixture may be more subtle.      

  The inset in Fig. \ref{fig11} shows the
nonzero polymer concentration    correlations on large
scales. Separation is now normalized by    colloid diameter, which
highlights the oscillatory correlations    due to imprinting of
colloid structural order on segmental    packing. The weakening of the
oscillatory features with increasing    polymer size is intuitively
expected and is also evident in the    dilute polymer limit.

\begin{figure}[ht]    \centerline{{\epsfysize=7.cm
\epsffile{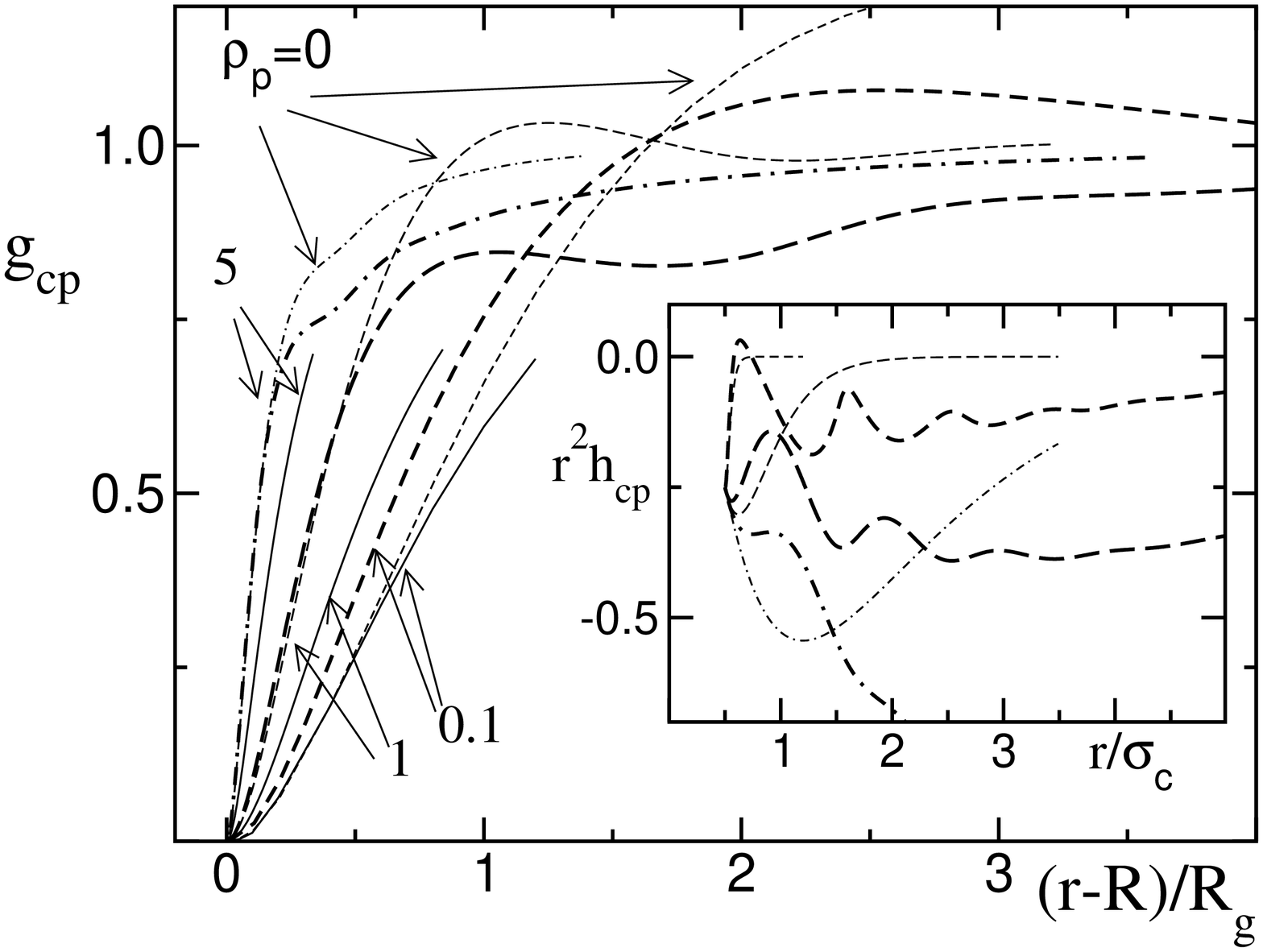}}}    \caption{ The main panel shows polymer
segment density profiles,    $g_{cp}$, versus rescaled distance,
$(r-R)/R_g$, as in Fig.    \protect\ref{fig11} but at colloid
concentration $\phi_c=0.20$;    size ratios as labeled and $c/c_s=0.8$
$(0)$ thick lines (thin    lines). Thin solid lines give the results at
these polymer    concentrations but at $\phi_c=0$. The inset shows the
curves at    $c/c_s=0.8$ and 0 in the form $r^2 h_{cp}(r)$ versus
$r/\sigma_c$    in order to exhibit the long-range
behaviour. \label{fig12} }    \end{figure}

Figure \ref{fig12} presents analogous results for a lower colloid volume
fraction where the pure hard spheres have very little solvation
shell structure. Our conclusions regarding the role of nonzero
polymer concentration are qualitatively the same as for Fig. \ref{fig11}:
the width of the local depletion layer is dominated by the
colloidal concentration fluctuation length scale. However, small
narrowing of the local depletion hole is now observable, and more
so as polymer size shrinks, due to weak polymer mesh formation. On
macromolecular distances, suppression of the (now weak) layering
behavior is again observed, and more so as the colloids increase    in
relative size.      

  The inset of Fig. \ref{fig12} shows the surface area
weighted correlated    part of $g_{pc}(r)$ as a function of separation
normalized now by    the colloid diameter. This plot again emphasizes
the imprinting of    oscillatory colloidal packing  on segmental
organization. However,    it  also demonstrates that there is still a
long range (power law    tail)  depletion layer component which
suppresses segment-colloid    contacts and is not screened away. The
oscillatory features    effectively "ride" this long range depletion
tail. As a subtle    consequence of the latter, even though the local
depletion layer    is relatively narrow on the $\sigma_{c}$ scale  for
$\phi_{c}=0.45$ , $c/c^{s}=0.8$ and $R_{g}/R=5$ (see inset of Fig.
\ref{fig11}), nevertheless the corresponding $g_{cc}(r)$ in Fig.
\ref{fig7}      looks like what one would expect for a weak, long
range attraction. This again emphasizes the existence of a long
range tail in $g_{pc}(r)$ connected to the $R_{g}$ scale. The
predicted scaling law in the limit of $R_g\gg R$ again bears this
out, as it gives for the large distance behaviour of the depletion
layer: 
\beq{eq15} g_{cp}(r\gg R) \to 1 - \frac{R}{r} \, A(\phi_c)
\; e^{-r/\xi_c} \fur R_g\gg R\; , \eeq 
where the parameters are
defined in \gls{eq13}{eq14}. Thus, the depletion layer exhibits a
power-law tail extending out to the collective correlation length
$\xi_c$, which increases upon approaching phase separation as seen
in the inset of Fig. \ref{fig12}. As argued below \gl{eq14}, we
expect this result to be qualitatively valid as long as the    polymer
mesh size is larger than the particle, $\xi > R$.        

\begin{figure}[ht]
\centerline{{\epsfysize=7.cm        \epsffile{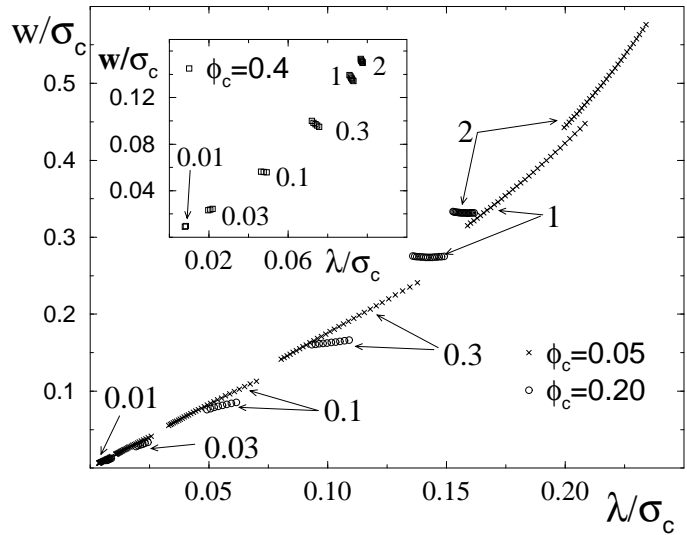}}}
\caption{ Local width of the polymer segment depletion layer as in
Fig. \protect\ref{fig5} versus $\lambda$ for two colloid
concentrations, $\phi_c=0.05$ and 0.20 and at  finite polymer
concentration. The curves are labeled with the size ratio
$R_g/(\sqrt2\sigma_c)$, and the polymer concentration increases up
to $c/c^*=1.5$ for $\phi_c=0.05$ and up to the spinodals for
$\phi_c=0.20$, respectively. $\lambda$ decreases with $c/c^*$.
The inset shows equivalent results for
$\phi_c=0.40$. \label{fig13} }    \end{figure}

The local
width of the polymer depletion layer at nonzero polymer
concentrations is shown in Fig. \ref{fig13}, and can be contrasted with
its dilute limit analog of Fig. \ref{fig5}. For low $\phi_c$, or for    small
polymer sizes such that $w\approx \xi$, the depletion width    $w$
correlates strongly with $\lambda$ from \gl{eq7}, the length    scale
over which the polymer segments rearrange. For the example    of
$\phi_c=0.05$ in Fig \ref{fig13}, both lengths decrease with
increasing $c/c^*$ and increase with $R_g/R$, yet saturating at
fractions of $R$ for $R_g/R\gg 1$. The primary finding for    slightly
higher colloid concentrations and/or polymer sizes such    that
$\xi\gg w$ is the relatively small influence of nonzero    polymer
concentration on the local depletion layer width. The    harsh
repulsive particle packing constraints overcome the    considerations
of polymer conformational entropy and the depletion    layer width $w$
becomes much smaller and almost independent of    $c/c^{*}$, while the
nonlocality length decreases with the polymer    correlation
length. While the case $\phi_c=0.4$ in Fig.    \ref{fig13} only shows
this behavior, for $\phi_c=0.2$ a crossover    between both trends
happens when the depletion layer becomes of    order of the relevant 
polymer length. Clearly, predicting the    depletion layer width at
finite concentrations becomes a difficult    quantitative problem in
general, and the accuracy of our PRISM    m-PY results need to be
tested.        

\begin{figure}[ht]    \centerline{{\epsfysize=7.cm
\epsffile{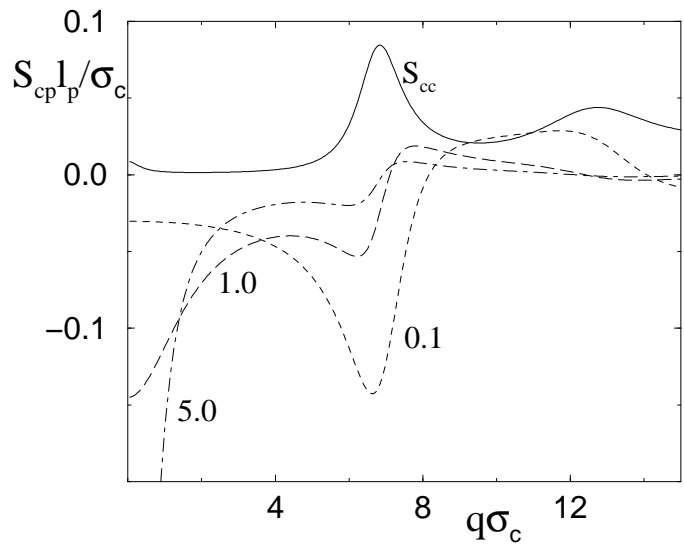}}}    \caption{Polymer-colloid (cross-term)
partial structure factors,    $\frac{l_p}{\sigma_c}\,S_{cp}(q)$, at
colloid concentration    $\phi_c=0.45$ and (large) constant distance
to the spinodal,    $c/c_s=0.3$, for the three size ratios
$R_g/R=0.1$, 1 and 5 as    labeled.
The    thin solid line gives the corresponding
(scaled) colloid structure    factor   $\frac{1}{30} S_{cc}(q)$ for
$R_g/R=5$. \label{fig14} }    \end{figure}

The Fourier space consequences of polymer-colloid
packing    correlations are probed via $S_{cp}(q)$, an example of
which is    shown in Fig. \ref{fig14}. The corresponding colloidal structure
factor    for the largest polymer case is also shown. There are three
distinct features. At small wavevectors, a peak is observed for    the
two larger polymer cases associated with the development of    long
wavelength concentration fluctuations which favor    polymer-colloid
demixing. It  also is present in the colloidal    scattering function
and is a precursor to phase separation since    the correlation length
associated with it is $\approx 2.5    \sigma_{c}$ which is larger than
even the dilute polymer solution    correlation length. On local
particle packing length scales, $q    \sigma_{c} \approx 7$,  a
negative, liquid-like packing feature is    observed associated with
anti-correlation of polymer and particle    positions on the colloidal
first neighbor shell scale. Its    amplitude increases as the polymers
get smaller in accord with the    reasoning leading to \gl{eq12s} for
$c\ll c^*$. On even smaller    (larger) length (wavevector) scales,
$S_{cp}(q)$ changes sign    implying positive excess polymer and
colloid density are now    favored.    

    \section{Polymer-polymer correlations}        

\subsection{Dilute polymers}      

\begin{figure}[ht]
\centerline{{\epsfysize=7.cm        \epsffile{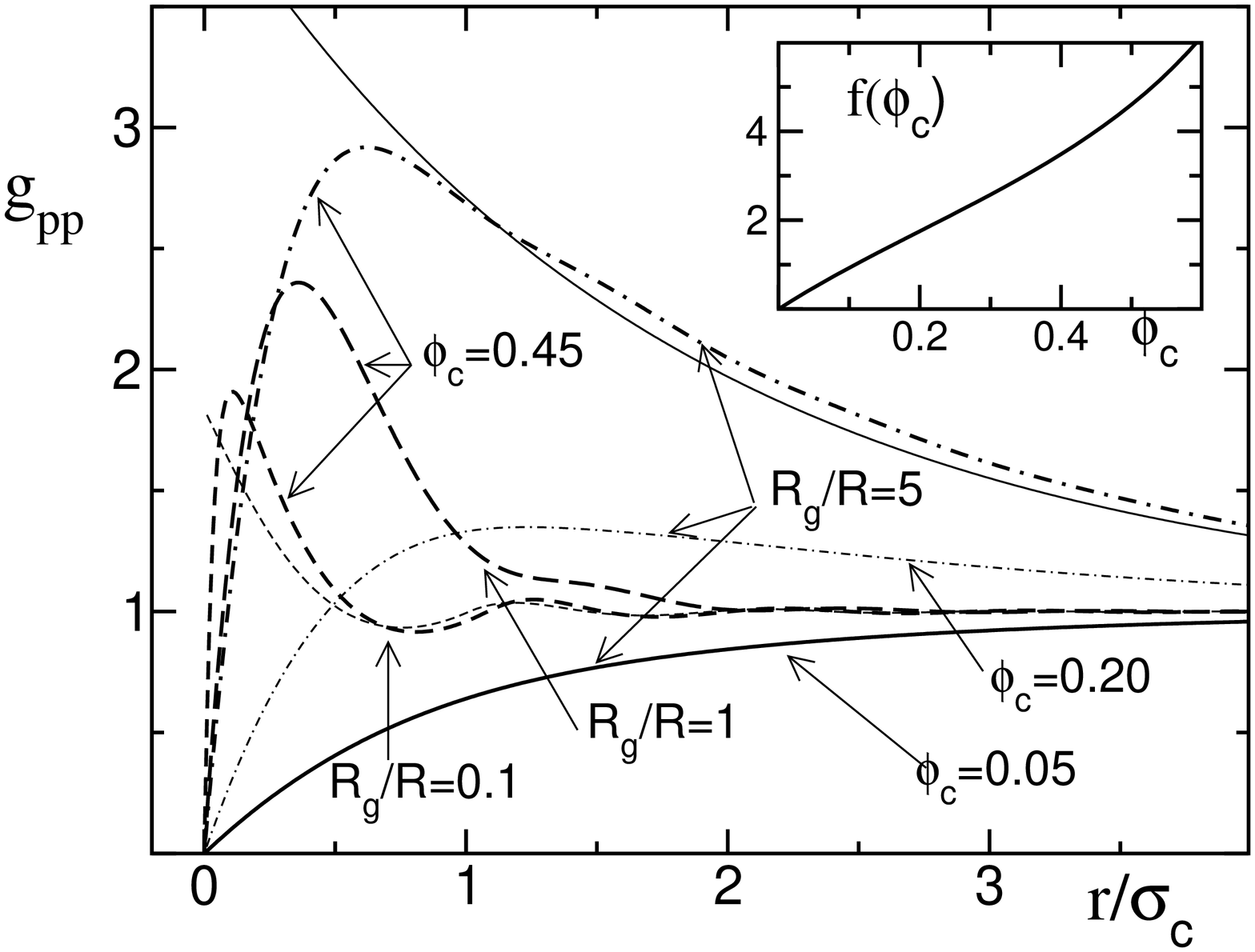}}}
\caption{ Polymer segment-polymer segment  pair correlation
function, $g_{pp}(r)$, for dilute polymers immersed in a hard
sphere solution; the colloid packing fractions $\phi_c$ and size
ratios $\xi_0$ are the same as in Fig. \protect\ref{fig4} and as
labeled.  The thin solid line presents the asymptote for large
polymers, Eq.  (\protect\ref{eq16}), evaluated for $R_g/R=5$ and
$\phi_c=0.45$, while the thin dashed line gives Eq.
(\protect\ref{eq15s}) at this $\phi_c$ which holds for $R_g\to0$.
The inset shows the intermolecular polymer segment contact value
$f(\phi_c)$, which determines the former asymptote. \label{fig15}
}\end{figure}

  Examples of
the interchain site-site correlations in the dilute    limit are shown
in Fig. \ref{fig15}.  The molecular polymer-polymer    second virial
coefficient is proportional to the integral of
$(1-g_{pp}(r))$.  In the absence of colloids, PRISM properly
recovers the classic $g_{pp}(r) < 1$ correlation hole behavior
characterized by the length scale $\xi_{0}$ that is indicative of
the ``soft-sphere'' character of the polymer coils. This
correlation hole is overcome at a distance related to the    depletion
layer of $g_{pc}(r)$ as seen in Fig. \ref{fig4}. With    increasing colloid
volume fraction and/or relative polymer size,    $g_{pp}(r)$ becomes
increasingly positive corresponding to a    colloid-mediated
attractive polymer-polymer second virial    coefficient \cite{17}. A peak
develops at finite $r$, and a mostly    structureless decay to the
random value of unity occurs with a    characteristic correlation
length which increases with $R_{g}$.   

     For small polymers at high
particle density, the segmental packing    displays an oscillatory
behavior due to efficient imprinting of    colloidal structure on the
polymer organization; see the    discussion below \gl{eq12s}. The
polymer coils can explore the    free volume homogeneously and thus
their correlations are    equivalent to the ones of the particles,
$g^{\rm HS}$, convoluted    with their (fixed) internal density
correlations, $P(r)=    (r-\sigma_c)^2\, (\sigma_c+r/2) \;
\Theta(\sigma_c-r)$ whose    Fourier transform is proportional to the
particle form factor    $P(q)$:
 \beq{eq15s} g_{pp}(r)=1 +
\frac{\pi\varrho_c/6}{(1-\phi_c)^2} \;\left[ P(r) + \varrho_c
\int\!\!d^3s\, P({\bf r}-{\bf s})\; (g^{\rm HS}(s)-1)\right] \fur
R_g\to0\; . \eeq 
This limiting result is included in Fig.
\ref{fig15} and explains the higher (than corresponding to the
average density $\varrho_p$) probability of two small coils to be
close at small distances.    

    In the $R_{g}/R \gg 1$ nanoparticle
limit, an asymptotic analytic    expression can be derived for
$g_{pp}(r)$ \cite{17}: 
\beq{eq16}    g_{pp}(r) \to 1 + ( f(\phi_c) - 1 ) \;
e^{-r/\xi_0}\; , \eeq
 where    $f(\phi_c)$ from \gl{eq14} is shown in
the inset of Fig.    \ref{fig15}. The transition from a depletion hole
to enhanced    packing ($g_{pp}(r)$ approaches unity from above)
occurs at    $\phi_{c} = 0.11$. For $\phi_{c} =0.5$, the density of
segments on    the second chain within a distance of $R_{g}$ from a
tagged    segment on the first chain is more than an order of
magnitude    larger than in the absence of colloids. The
colloid-induced    enhancement of polymer-polymer contacts is a
phenomenon not    treated by effective-potential models. It depends
systematically    on $R_{g}/R$ and $\phi_{c}$, and implies a tendency
for correlated    polymer clustering. The latter has strong
thermodynamic    consequences, and plays an essential role in the
novel predictions    of PRISM theory for the fluid-fluid demixing
transition    \cite{16,17,32}.            

\subsection{Concentrated polymers}    

\begin{figure}[ht]    \centerline{{\epsfysize=7.cm
\epsffile{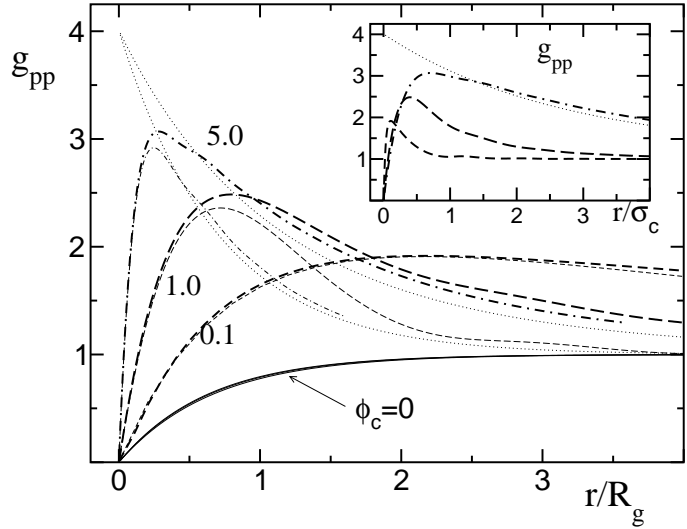}}}    
\caption{ Polymer segment-polymer
segment pair correlation    functions, $g_{pp}(r)$, versus radial
distance in units of    $R_g$  for the size ratios, $R_g/R=0.1$ (short
dashes), 1.0 (long    dashes) and 5.0 (dot-dashed). As in
Fig. \protect\ref{fig11} the    colloid concentration is
$\phi_c=0.45$, and the polymer    concentration relative to the
spinodal is constant, $c/c_s=0.8$    (bold) and $0$ (thin). The thin
solid (almost overlapping) lines    give the results for $c/c_s=0.8$
and vanishing colloid    concentrations, $\phi_c=0$. Dotted lines
indicate the asymptote    for $R_g/R\to\infty$ evaluated for the two
curves at $R_g/R=5$.    The inset shows the finite density results
(thick lines) and the    large $R_g$-asymptote from the main panel as
a function of    $r/\sigma_c$ (thin dashes).    
      \label{fig16} }    \end{figure}

    Examples of the influence of nonzero polymer
concentration on    polymer segment-segment pair correlations are
given in Fig. \ref{fig16}    for a high colloid volume fraction. The
analogous results for    vanishing polymer or colloid concentration
are also presented for    comparison. As found for the polymer-colloid
pair correlations,    the most striking aspect of Fig. \ref{fig16} is the
relative    insensitivity of the segmental correlations to polymer
concentration at the rather low  values of $c/c^{*}$ which
characterize the homogeneous one-phase regime.  For $r \ll R_{g}$,
a correlation hole is present which is only slightly narrower than
in the $c\to 0$, dilute polymer limit. The hole caused by the
mutual soft repulsion of the coils crosses over to an enhanced
segmental packing correlation ($g_{pp} >1$) on a length scale
comparable to $R_g$ ($\sigma_c$) for small (large) polymer radii.
Small increases of intensity and location of the latter feature
occur at nonzero polymer concentration, although the distance of
decay to the ultimate random value of unity is clearly enhanced    due
to pre-transitional long wavelength composition fluctuations.    The
$R_g\gg R$-scaling scaling law corresponding to    \gls{eq13}{eq15},
 is given by 
\beq{eq16s}     g_{pp}(r) \to 1    + \; \frac 1b\;
\frac{R}{r}\; \left( e^{-r/\xi_c} - e^{-r/\xi_0}    \right) \fur R_g
\gg R\; , \eeq 
This result demonstrates that long    wavelength
correlations can have an amplitude proportional to the    inverse of
the small parameter, $b\sim (c/c^*)(R/R_g)$, which    quantifies the
fraction of displaced segments. The dotted lines    show that the
asymptotically large polymer limit is rather closely    approached
already when $R_{g}/R = 5$, and that adding polymer    dominantly
increases the collective correlation length,  $\xi_c$    of \gl{eq14},
because it drives the system closer to  the    spinodal demixing
instability.      

  The inset of Fig. \ref{fig16} shows the nonzero polymer
concentration    results on an expanded scale with intersegment
separation    nondimensionalized by colloid diameter. A weak
oscillatory feature    is present for the smallest polymer, indicating
that they again    fill the voids between the particles rather
homogeneously so that    \gl{eq15s} could be generalizable to finite
densities.   

\begin{figure}[ht]    \centerline{{\epsfysize=7.cm
\epsffile{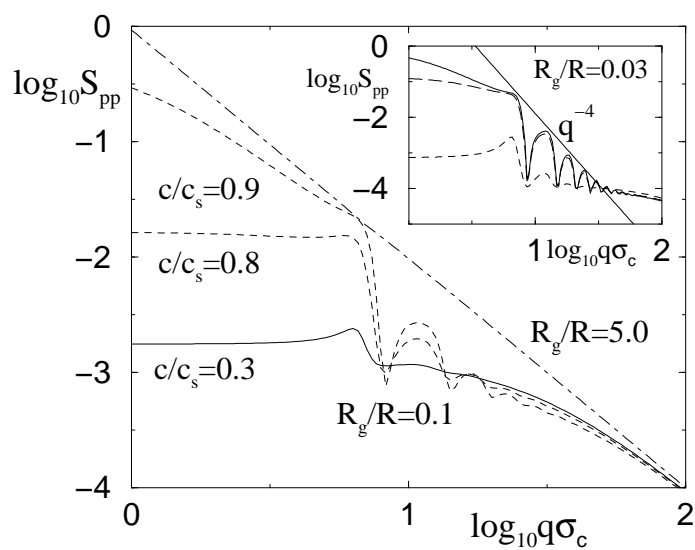}}}    \caption{Double-logarithmic plot of the
collective polymer    structure factor, $S_{pp}(q)$ in units of
$(\frac{\sigma_c}{l_p})^2$,
for colloid
packing fraction $\phi_c=0.45$, constant distance to the spinodal,
$c/c_s=0.8$ (thick lines) and two size ratios, $R_g/R=0.1$    (dashed)
and 5 (dash-dotted), as labeled. For the smaller    polymer, the
curves for $c/c_s=0.3$ (thin solid) and $c/c_s=0.9$    (thin dashes)
are also shown. The inset shows $S_{pp}$ for size    ratio
$R_g/R=0.03$ and the polymer concentrations $c/c_s=0.3$    (short
dashes), 0.6 (long dashes) and 0.8 (solid), while a thin    solid line
corresponds to the Porod-scattering law $q^{-4}$.    \label{fig17} }
\end{figure}

     An example of the polymer-polymer structure factor
is given in    Fig. \ref{fig17} in a log-log format. Results are shown for a
high    colloid volume fraction and a large and small value of
$R_{g}/R$.    For the larger polymer case, and even if $R_{g} = R$
(not shown),    the scattering function is very nearly a simple
Lorentzian, $S_{pp}(q)=\frac{(\xi_c/l_p)^2}{1+(q\xi_c)^2}$
 as    follows from the large $R_{g}$ asymptotic behavior,
\gl{eq16s}.   However, for    small
enough polymers such as the $R_{g}/R = 0.1$ case shown, the    polymer
scattering pattern displays distinctive regimes as a    function of
polymer concentration and wavevector. For $q R_{g}    >1$,
intramolecular single chain correlations are probed and the
Gaussian chain power law of $q^{-2}$ is found. For $q R_{g} < 1$,
but $q \sigma_{c} > 7$, oscillatory features emerge with    increasing
polymer concentration. As discussed in section 3.2, this    behavior is a
consequence of the simple fact that polymer density    must vanish
where the colloids are, and that in the free volume    space, where
the polymers are located, they are distributed in an    almost
homogeneous manner on scales larger than the
radius-of-gyration. Therefore, the relative amplitude, and maxima
and minima, should be well described by the classic form factor of
a homogeneous sphere:      $P(q) = \frac{3}{(qR)^{4}} [ \cos{qR} -
\sin{qR}/qR ]^{2}$. This is    indeed what we find, and the classic
$q^{-4}$ Porod law    appropriate for scattering from a sharp
interface between two    phases is observed with increasing polymer
concentration, and at    smaller values of $c/c_{s}$ as $R/R_{g}$
increases (see inset).     

   For other situations and parameter
values we find (not shown)    behavior similar to that shown in Fig.
\ref{fig17}.
 For example, at a    relatively high fixed value of $c/c^{*}$ for
$R_{g} \ll  R$, a    very similar sequence of scattering profiles is
found as the    spinodal is approached by increasing $\phi_{c}$.

\section{Discussion}    

    We close with a brief discussion of the
limitations of the present    version of PRISM/m-PY theory, and future
opportunities for    generalization and applications.  

      For the
athermal polymer-particle mixture problem, there are    presently five
primary limitations we can identify.  (1) Use of    the PY closure for
colloid-colloid direct correlations can break    down when the
depletion attraction becomes much stronger than the    thermal energy
kT \cite{16,52,57}. This is most relevant when $R\gg R_g$    and $\phi_c$ is
small, and may play an important role in the    prediction of complete
miscibility (in the spinodal sense) at    small colloid volume
fractions seen in Fig. \ref{fig6}. (2) The thread    model is not appropriate
at high (melt-like) polymer densities, nor    for persistent chains
mixed with relatively small particles,    situations for which the
specific monomer size and/or polymer    backbone stiffness become
relevant. (3) The use of a Gaussian    polymer structure factor for
athermal solution conditions instead    of the more open self-avoiding
walk model is expected to    overestimate depletion effects, and more
so as $R_g/R$ increases.    (4) A fully self-consistent treatment of
single chain correlations    which accounts for both screening of
intramolecular excluded    volume interactions by polymer
concentration fluctuations \cite{65},    and possible colloid-induced
conformational changes, has not been    carried out.  The latter
effect is expected to become more    important as $R_g/R$ increases
and polymers must "wrap around" the    particles.  (5) Our results for
fluid-fluid phase separation are    presently at the spinodal
instability (plus critical point) level;    numerical construction of
full binodal coexistence curves remains    to be achieved.  Such work
is in progress.    

    Both new experiments and computer simulations
would be very    valuable to test our predictions. Small and wide
angle scattering    experiments to extract the colloid-colloid, and
polymer-polymer,    structure factors can be directly compared with
the theoretical    results in wavevector space. Alternatively, in the
colloidal    regime real space video microscopy measurements could
test our    radial distribution predictions for
$g_{cc}(r)$. Systematic phase    diagram studies in the true
nanoparticle regime over a wide range    of particle volume fractions
could further test our novel    predictions for fluid-fluid phase
separation and the location of    critical points. Computer
simulations which include    polymer-polymer repulsive forces should
be feasible in the    nanoparticle regime, and would provide detailed
information    concerning both structure and miscibility. Simulations
of the    elementary 1 and 2 particle (or polymer) problems appear to
not be    available and would be the best model systems to first
investigate    and compare with theory.

     Regarding future
extensions of the integral equation theory there    are multiple
possibilties which present variable degrees of    technical and
conceptual challenges. First, either analytically,    or with modest
numerical effort, our results for 1 and 2 particles    in a polymer
solution can be generalized to non-Gaussian coil    architectures
characterized by arbitrary fractal dimensions    \cite{47,48}. This is
especially relevant to experiments which probe    the second virial
coefficient of nanoparticles (e.g, proteins    \cite{56}) which are
sensitive to polymer conformational statistics    \cite{27}.    

    Second,
at nonzero concentrations of both species our use of the    Baxter
factorization method to (nearly) solve the coupled integral
equations is no longer applicable for polymers that are not
described by a Gaussian structure factor. Hence, a fully numerical
approach must be employed. This will be most tractable within the
"thread" level description where the polymer monomer excluded
volume diameter is shrunk to zero. Variable polymer structures
(e.g., rods, stars) can be modelled by adopting the appropriate
single macromolecule structure factor. Semiflexible chains can be
modelled by introducing the chain persistence length scale, and    the
crossover of $\omega(q)$ from rigid rod like to random coil    like
with decreasing wavevector. Modification of the detailed form    of
the m-PY closure is expected to be required to properly reflect
polymer conformational statistics on length scales smaller than    the
nonlocality length $\lambda$. Based on a numerical approach,
generalization to treat nonspherical particles, or spheres with
heterogeneous surfaces \cite{66}, is natural within the site
representation of RISM theory as long as strong orientational
correlations or liquid crystal formation are not relevant \cite{21}.

In principle, one can address the role of non hard core forces,
such as van der Waals attractions or Coulomb interactions \cite{67},
between polymer and/or colloidal species, including the question    of
variable solvent quality for the polymer chain. However, this    can
be a difficult task within integral equation theory when such
variable strength and spatial range interactions can induce
significant changes in mixture structure. For example, if strong
attractions (or non contact repulsions) exist between polymer
segments and particles, then collective phenomena akin to
wetting/adsorption (dewetting/drying) can occur which are    generally
not properly captured with standard closure    approximations
\cite{68}. However, the availability of field theoretic    results for the
simplest realization of such problems \cite{6,27,28,29}    may again provide
critical guidance for the development of new    closure
approximations. The question of conformational    perturbations of the
polymer may also become more important,    especially if the polymer
experiences strong attractions with the    colloids or
nanoparticles. Application of existing self-consistent    schemes
\cite{6,26,65,69}, including hybrid PRISM plus field theoretic    or Monte
Carlo simulation approaches which have successfully    treated
intrapolymer excluded volume \cite{65}, are then required. Of    course,
the treatment of Coulomb forces remains a major challenge    even for
spherical colloids \cite{67}, although progress has been    recently made
within the integral equation framework for charged    spheres and/or
polyelectrolytes \cite{65,70}.   

     Finally, high density macromolecular
systems are relevant to    materials such as (nano-) particle filled
polymer melts. Here a    thread description is not appropriate, but
the basic m-PY idea    should again be applicable to account for the
local (chemically    specific) perturbation of polymer chains near a
particle surface.    An explicit treatment of the nonzero monomer
diameter and other    local chain structural features is required,
which should be    numerically feasible if the particles are of
nanoscopic    dimensions.        

\section{Acknowledgments}
\label{acknowledgements}    

    We acknowledge helpful discussions
with L. Belloni, A.P. Chatterjee,    Y.L. Chen, S. Egelhaaf,
E. Eisenriegler, A. Johner, A. Moussaid, W. Poon, P. Pusey,    M. Schmidt and
C. Zukoski. M.F. was supported by the Deutsche
Forschungsgemeinschaft under Grant No. Fu 309/3 and through the    SFB
563. K.S.S. was supported by the U.S. Department of Energy    grant
number DEFG02-91ER45439 through the UIUC Materials Research
Laboratory.

\newpage            

\end{document}